\documentclass[11pt,a4paper]{article}

\usepackage{graphicx}
\usepackage{amssymb}

\usepackage{booktabs}

\usepackage[detect-all]{siunitx}

\usepackage{extsizes}
\usepackage[version=3]{mhchem}
\usepackage{graphicx} 
\usepackage{lastpage}
\usepackage{float}
\usepackage{fancyhdr}
\usepackage{fnpos}
\usepackage[english]{babel}
\usepackage{array}
\usepackage{droidsans}
\usepackage{charter}
\usepackage[T1]{fontenc}
\usepackage[usenames,dvipsnames]{xcolor}
\usepackage{setspace}
\usepackage[compact]{titlesec}
\usepackage[hidelinks]{hyperref}
\usepackage{subfigure}
\usepackage{indentfirst}

\usepackage{cite}

\usepackage{geometry}
\usepackage{authblk}

\graphicspath{{figs/}}

\title{Belousov-Zhabotinsky reaction in liquid marbles}

\date{}                                           

\author[1]{Claire Fullarton}
\author[1]{Thomas C.\ Draper}
\author[1]{Neil Phillips}
\author[1,2]{\\Ben P.\ J.\ de Lacy Costello}
\author[1]{Andrew Adamatzky$^{\ast}$}

\affil[1]{\small{Unconventional Computing Laboratory, University of the West of England, Bristol, UK}}
\affil[2]{\small{Institute of Biosensing Technology, Centre for Research in Biosciences, University of the West of England, Bristol BS16 1QY, UK}}

\newcommand\blfootnote[1]{%
  \begingroup
  \renewcommand\thefootnote{}\footnote{#1}%
  \addtocounter{footnote}{-1}%
  \endgroup
}

\providecommand{\keywords}[1]{\textbf{\textit{Keywords:}} #1}

\begin{document}

\maketitle

\begin{abstract}

In Belousov-Zhabotinksy (BZ) type reactions, chemical oxidation waves can be observed and the oscillations of these waves can be exploited to produce reaction-diffusion processors. This paper reports on a new method of encapsulating BZ solution droplets in a powder coating of either polyethylene (PE) or polytetrafluoroethylene (PTFE), to produce BZ liquid marbles (LMs). The BZ LMs have solid--liquid interfaces in contrast to the previously reported BZ encapsulation systems used to study pattern formations and wave transfers, such as BZ emulsions and BZ vesicles. All these systems are important to study as they can be seen as analogues for information transfer in excitable systems in nature. Due to the powder coating of the LMs, LMs do not wet the underlying substrate, making the LMs mobile, enabling them to be easily arranged in arrays. The preparation of complex LMs, containing an acidic oscillating solution, proves the resilience of LMs and their application in transporting a wide range of chemical cargoes and in playing a mediating role in chemical reactions. PTFE-coated LMs were harder to prepare and not as robust as PE-coated LMs, therefore oscillation studies of BZ LMs positioned in arrays only focused on PE-coated LMs. Sporadic transfer of excitation waves was observed between LMs placed in close proximity to each other in ordered and disordered arrays, suggesting the transfer of liquid species possibly arises from contact between imperfectly coated areas at the LM--LM interface or capillary action, where solution is actively transported to the marble surface through the coating. Propagation pathways of the excitation waves in both the disordered and ordered arrays of BZ LMs are reported. These results lay the foundations for future studies on information transmission and processing arrays of BZ LMs, in addition to observing BZ wave propagation in a LM (which acts as an unstirred reactor vessel) and the effect of other physical stimuli such as heat, light and chemical environments on the dynamics of excitation in arrays of BZ LMs.

\end{abstract}

\noindent \keywords{Liquid marbles, Belousov-Zhabotinsky (BZ) reaction,  wave propagation}

\blfootnote{\textit{$^{\ast}$~Corresponding author. Email: Andrew.Adamatzky@uwe.ac.uk}}

\section*{Introduction}

Belousov--Zhabotinsky (BZ) type reactions involve the oxidation of an organic substrate (typically malonic acid) by bromate ions in the presence of an acid (typically sulphuric acid) and a one electron transfer metal ion redox catalyst (e.g. cerium, ferroin, and / or the light sensitive ruthenium complex [tris(2,2'-bipyridyl)-ruthenium(II)] \ce{[Ru(bpy)3]^2+}~\cite{belousov1959periodic,zhabotinsky1964periodic,Kuhnert1986}.

The BZ reaction involves three major reactions; firstly the reduction of bromide ions by bromate ions, secondly the autocatalytic species, the bromous acid (\ce{HBrO2}), oxidises the reduced form of the catalyst ferroin (Fe(II)) to ferriin (Fe(III)), and thirdly, when the Fe(III) reaches a high enough concentration, this initiates the reaction of the organic substrate (in this case, malonic acid) and its brominated derivative (bromomalonic acid) to yield the reduced catalyst, ferroin and bromide ions (reaction inhibitor), which initiates the first reaction again. The changes in oxidation state of the catalyst (Fe(II) / Fe(III)), result in a colour change from red to blue. The intermediate \ce{HBrO2} is the reaction activator (autocatalytic species), the diffusion of which enables the propagation of waves through the media.  

These reactions have proved to be potential media for developing future and emergent computing devices based on the interaction of chemical wave-fragments. A substantial number of theoretical studies and experimental prototypes of computing devices have been implemented using this media; image processors and memory devices~\cite{Kuhnert1986, Kuhnert1989, Kaminaga2006}, logical gates implemented in geometrically constrained BZ media~\cite{Steinbock1996, Sielewiesiuk2001}, approximation of the shortest path of excitation waves~\cite{Steinbock1995, Rambidi2001, Adamatzky2002}, information coding using the frequency of oscillations~\cite{Gorecki2014}, onboard controllers for robots~\cite{Adamatzky2004, Yokoi2004, VazquezOtero2014}, chemical diodes~\cite{Igarashi2011}, neuromorphic architectures~\cite{Gorecki2009, Stovold2012, Gentili2012, takigawa2011dendritic,  Gruenert2015} and associated memory~\cite{stovold2016reaction,stovold2017associative}, wave-based counters~\cite{gorecki2003chemical} and other information processors~\cite{DBLP:journals/ijuc/YoshikawaMIYIGG09, escuela2014symbol, Gruenert2015, gorecki2015chemical}. 

To provide insights into these processes and the use of the BZ media for the applications above, studies have focused either on thin films of media or on confining the BZ reaction in small droplets in microfluidic devices~\cite{chang2016fabricating, Gruenert2015, Steinbock1998, Gizynski2017b, Guzowski2016, Kitahata2002, Kitahata2012, Litschel2017, Suematsu2016, Stricker2017, Torbensen2017, Torbensen2017b}, vesicles~\cite{Tamate2017, Tamate2017b, Hu2016, Stockmann2015, deSouza2014}, emulsions~\cite{Wang2016}, BZ-AOT emulsions~\cite{Kaminaga2006, Rossi2008, Bansagi2011, Vanag2001, Vanag2001b, Vanag2002, Vanag2003, McIlwaine2009, Gong2003, Cherkashin2017}, cation exchange resins~\cite{Maselko1989, Taylor2015, Nishiyama1994, Nishiyama1995, Aihara2001, Kuze2018, Totz2018} and 3D printed structures~\cite{king2014interdroplet,king2015excitability,chang2016fabricating}.

Liquid marbles (LMs), first reported by Aussillous and Qu\'{e}r\'{e}~\cite{Aussillous2001}, provide a means of moving droplets across a surface without wetting the underlying substrate, which avoids surface contamination problems and allows fast displacement and easy manoeuvrability of the droplets. In the context of using LMs as a method for encapsulating the BZ reaction, they avoid the use of complex microfluidic systems and allow BZ media to be probed in a solid--liquid system rather than in a liquid--liquid system e.g. BZ media in vesicles and emulsions. Properties of LMs can be tailored for a variety applications by altering the encapsulated liquid and / or the powder coating~\cite{Bormashenko2017, Bormashenko2012, Fujii2016, McHale2015, Ooi2015, rychecky2017spheroid}. This tailoring can enable LMs to be manipulated using electric and magnetic fields, as well as being able to be mechanically manipulated. Therefore, LMs can be forced to merge together (through collisions or field effects), split into daughter LMs (if there is a sufficient amount of coating on the daughter LMs to not wet the underlying substrate), opened, closed, easily modified in terms of adding and removing liquid from a pre-made LM, as well as being permeable to gas, due to the powder coating~\cite{Draper2017, Aussillous2006, Bormashenko2011,  Bormashenko2011a, Bormashenko2015b, Bhosale2008, Zhao2010}.  

This paper reports on the preparation of acidic LMs using BZ solution as the liquid encapsulated in a polymer powder coating (polyethylene (PE) or polytetrafluoroethylene (PTFE)). The BZ solution inside the LM was monitored to observe whether pattern formation occurred and if waves propagated inside the LM.  Arrays of BZ solution LMs prepared, assessed whether wave transfers occurred between LMs in close proximity as well as observing the collective behaviour of the oscillating system. These experiments lay the foundations for developing unconventional computing devices using LMs as a means of encapsulating BZ solution, in a system which can be effortlessly reconfigured into various different architectures.

\section*{Experimental}

The BZ reaction (ferroin-catalysed / malonic acid BZ reaction) studied was prepared using the method reported by Field~\cite{Field1979}, omitting the surfactant Triton X. 18~M Sulphuric acid \ce{H2SO4} (Fischer Scientific, CAS 7664-93-9), sodium bromate \ce{NaBrO3} (Sigma Aldrich, CAS 7789-38-0), malonic acid \ce{CH2(COOH)2} (Sigma Aldrich, CAS 141-82-2), sodium bromide \ce{NaBr} (Sigma Aldrich, CAS 7647-15-6) and 0.025~M tris-(1,10-phenanthroline) iron(II) sulphate (ferroin indicator, Sigma Aldrich - Honeywell Fluka, CAS 14634-91-4) were used as received. Coatings for LMs, ultra high density polyethylene (PE) (Sigma Aldrich, CAS 9002-88-4, Product Code 1002018483) and polytetrafluoroethylene (PTFE) (Alfa Aesar, CAS 9002-84-0, Product Code 44184) were used as received, with particle sizes \SI{100}{\micro\m} and \SI{6}{\micro\m}--\SI{10}{\micro\m} respectively.

\ce{H2SO4} (2~ml) was added to deionised water (67~ml), to produce 0.5~M \ce{H2SO4}, \ce{NaBrO3} (5~g) was added to the this to yield 70~ml stock solution, containing 0.48~M \ce{NaBrO3}. Stock solutions of 1~M malonic acid and 1~M NaBr were prepared by dissolving 1~g in 10~ml of deionised water.

In a 50~ml beaker, 0.5~ml of 1~M malonic acid was added to 3~ml of the acidic \ce{NaBrO3} solution. 0.25~ml of 1~M NaBr was then added to the beaker, which produced bromine. The reaction was left, until a clear colourless solution remained (ca. 5~mins) before adding 0.5~ml of 0.025~M ferroin indicator to the beaker. 

BZ LMs were prepared by pipetting droplets of BZ reaction mixture (50 and \SI{100}{\micro\litre}), which was already oscillating, on to a powder bed of either polyethylene (PE) or polytetrafluoroethylene (PTFE) in a weighing boat, releasing the droplet ca. 5~mm from the top of the powder bed. The BZ droplet was rolled on the powder bed for 10~s to produce a LM. The single BZ LMs were then transferred on to a cool white LED housed in a black plastic box (single 5~mm diameter cool white LED 5000--8300K, powered by a standard 9V battery) to highlight the oscillating reaction inside the marble and enable the observation of travelling wave-fronts through the LM coating. Disordered arrays of LMs were  prepared by transferring the pre-made LMs into a Petri dish. Ordered LM arrays were prepared by rolling LMs onto a 4 x 4 plastic polypropylene template. Both BZ arrays were then illuminated by using an LED light underneath the Petri dish.  The BZ reaction in LMs was recorded using a USB microscope with magnification $\times 5$.

\section*{Results}

Initially, to investigate the feasibility of preparing stable LMs using the acidic BZ media, powder coatings of PE and PTFE were studied. The rationales behind using PE and PTFE were these polymers have previously been demonstrated as suitable coatings for LMs, are relatively inert to acidic solutions, provide a comparison between LM coating particle sizes and appear translucent when illuminated with an LED~\cite{Fullarton2018}. The latter meant it was possible to observe the oxidation waves through both PE and PTFE coatings. The difference in particle size of the coatings, varied the particle spacing on the surface of the LM, which therefore varied, to different degrees, the observations of the oscillating BZ solution within the LM. Studies focused on observing the colour changes and oxidation waves of the BZ media in single LMs and then observing the behaviour of oxidation waves in disordered and ordered arrays of LMs. The concentration of BZ solution used to make the LMs, exhibited oscillating behaviour. When the BZ solution prepared was left unstirred in a thin film in a Petri dish, pattern formation occurred. The full videos of all single LMs can be found in the ESI.

\begin{figure}[!tbp]
\centering
    \subfigure[1s]{\includegraphics[width=0.3\textwidth]{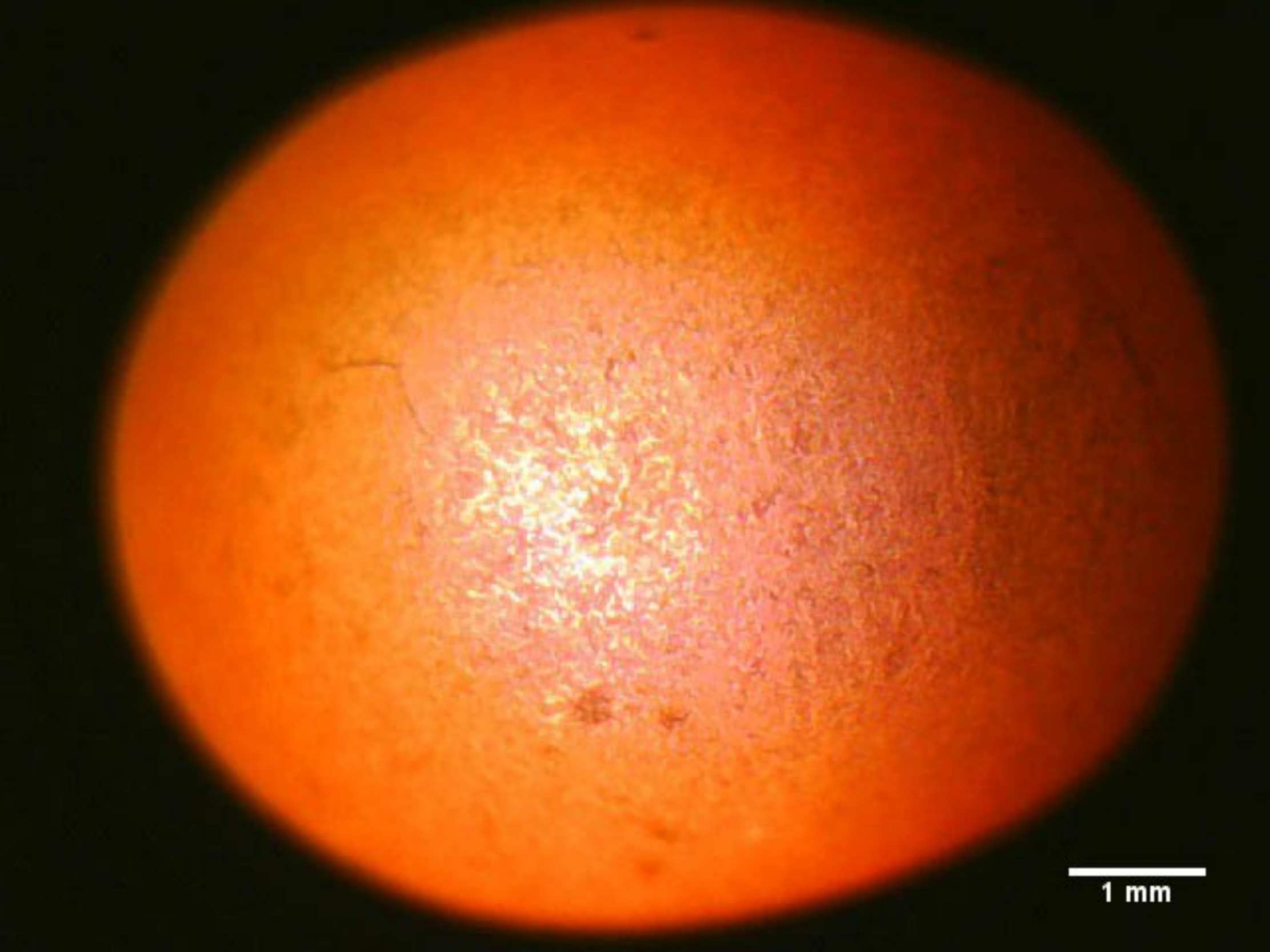}}
    \subfigure[691s]{\includegraphics[width=0.3\textwidth]{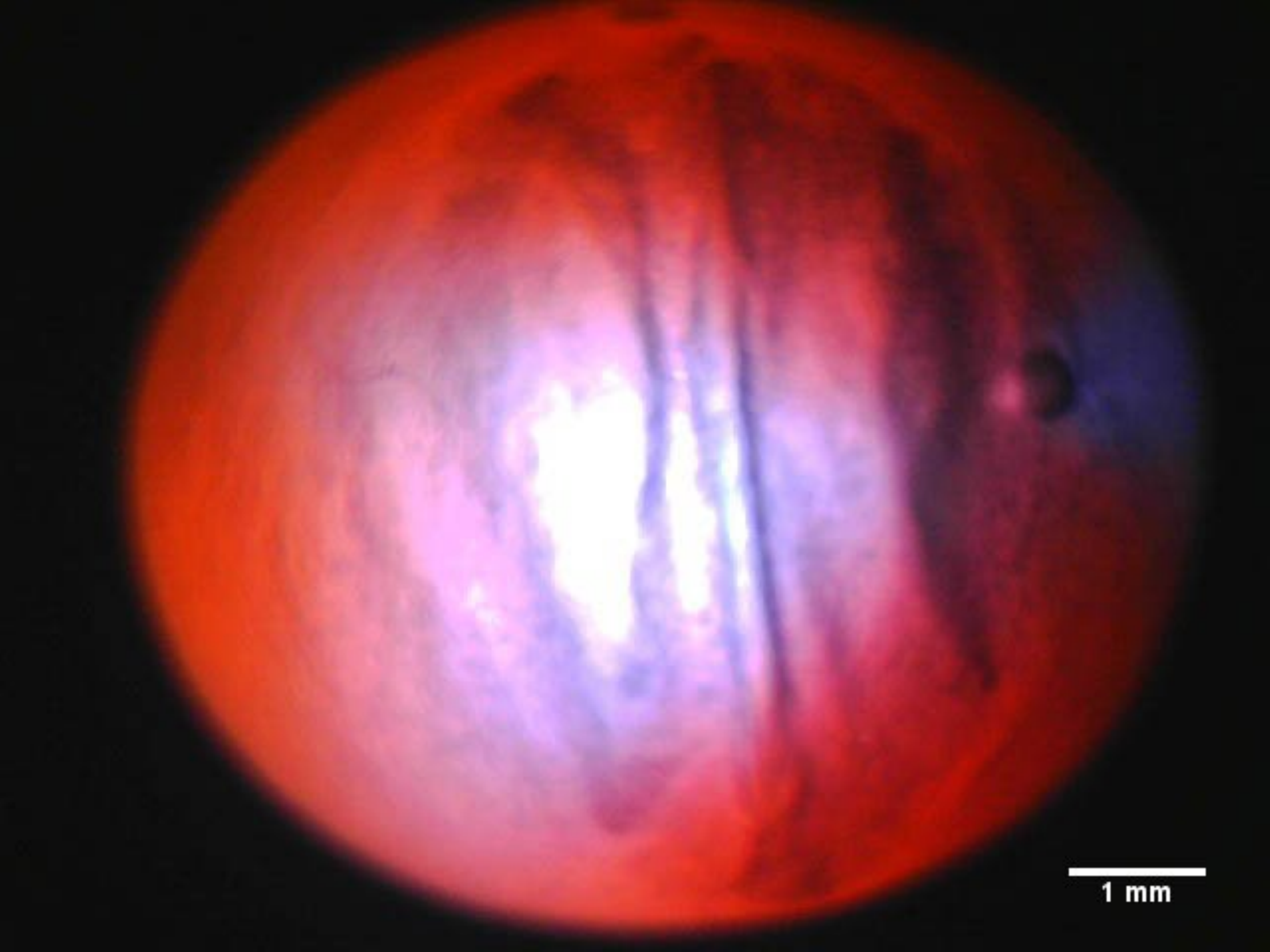}}
    \subfigure[697s]{\includegraphics[width=0.3\textwidth]{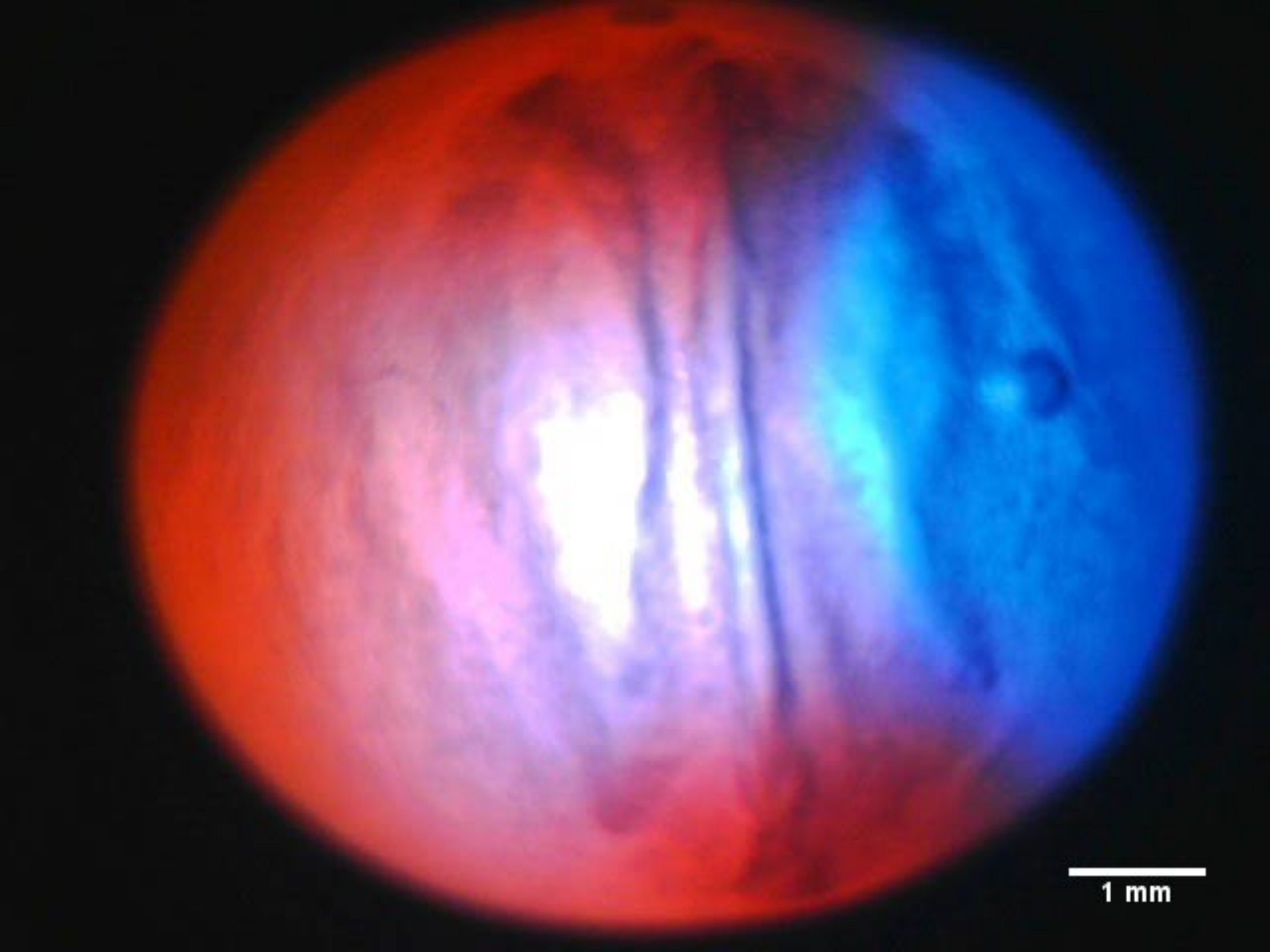}}
    \subfigure[703s]{\includegraphics[width=0.3\textwidth]{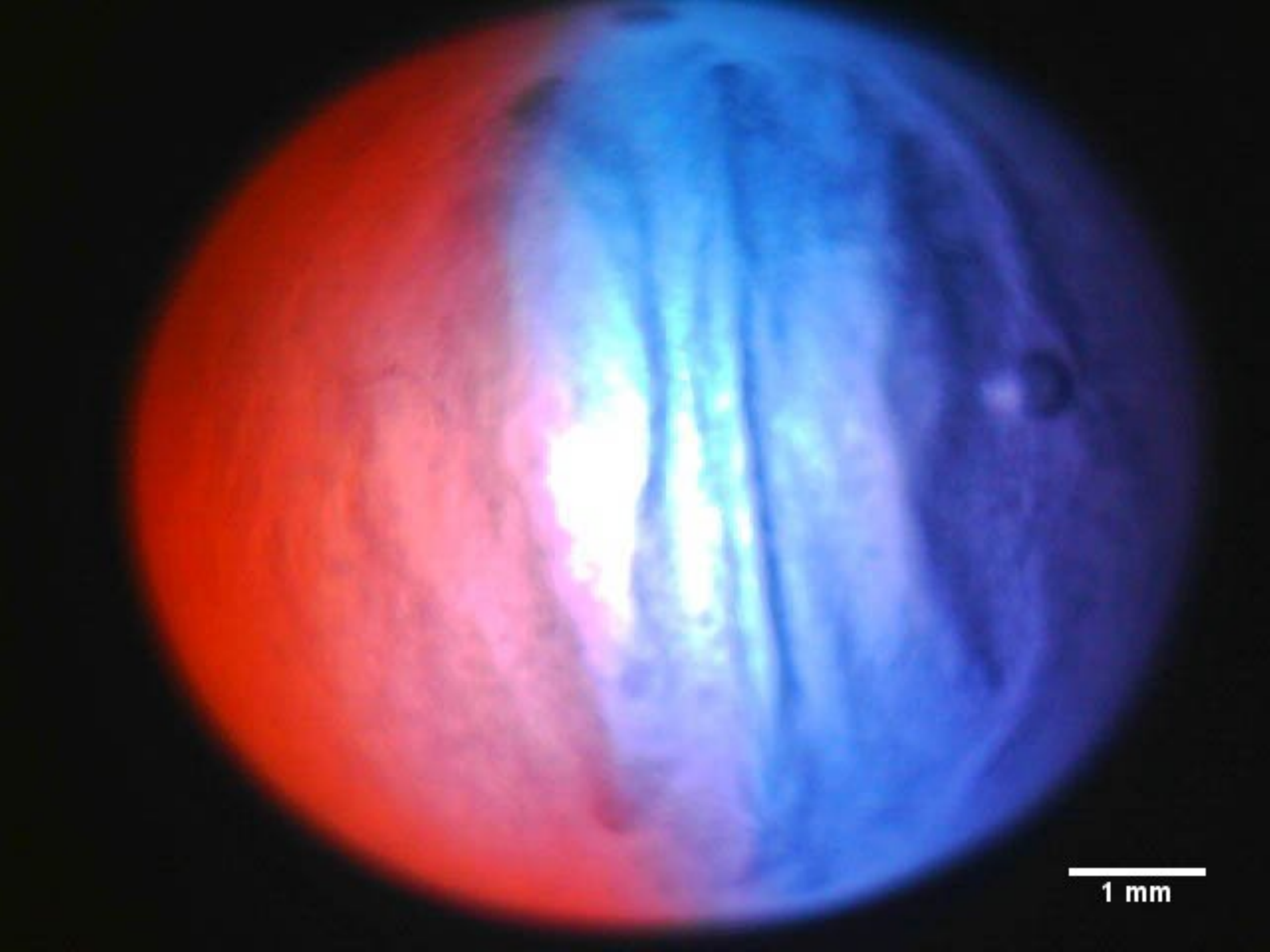}}
    \subfigure[709s]{\includegraphics[width=0.3\textwidth]{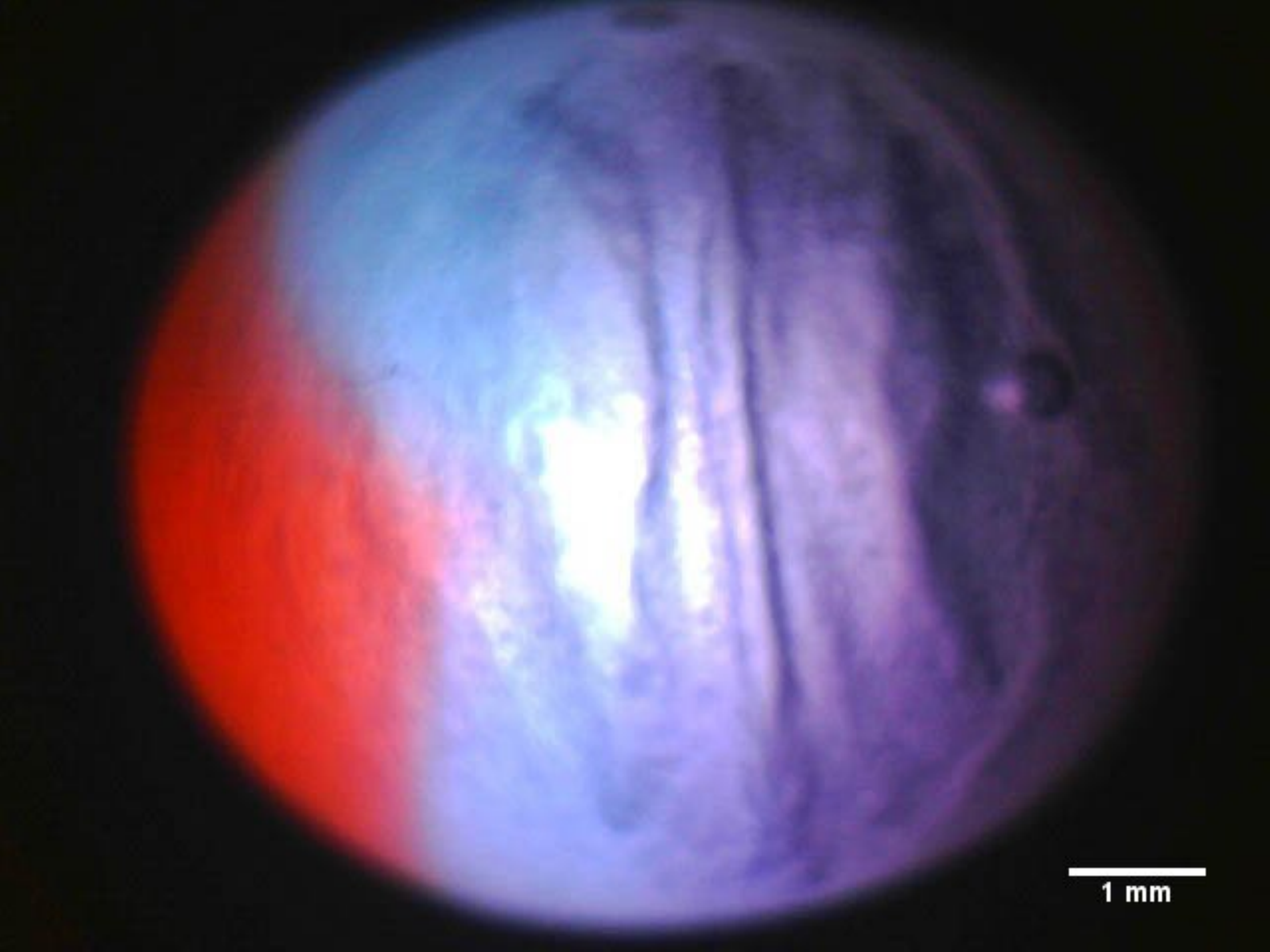}}
    \subfigure[716s]{\includegraphics[width=0.3\textwidth]{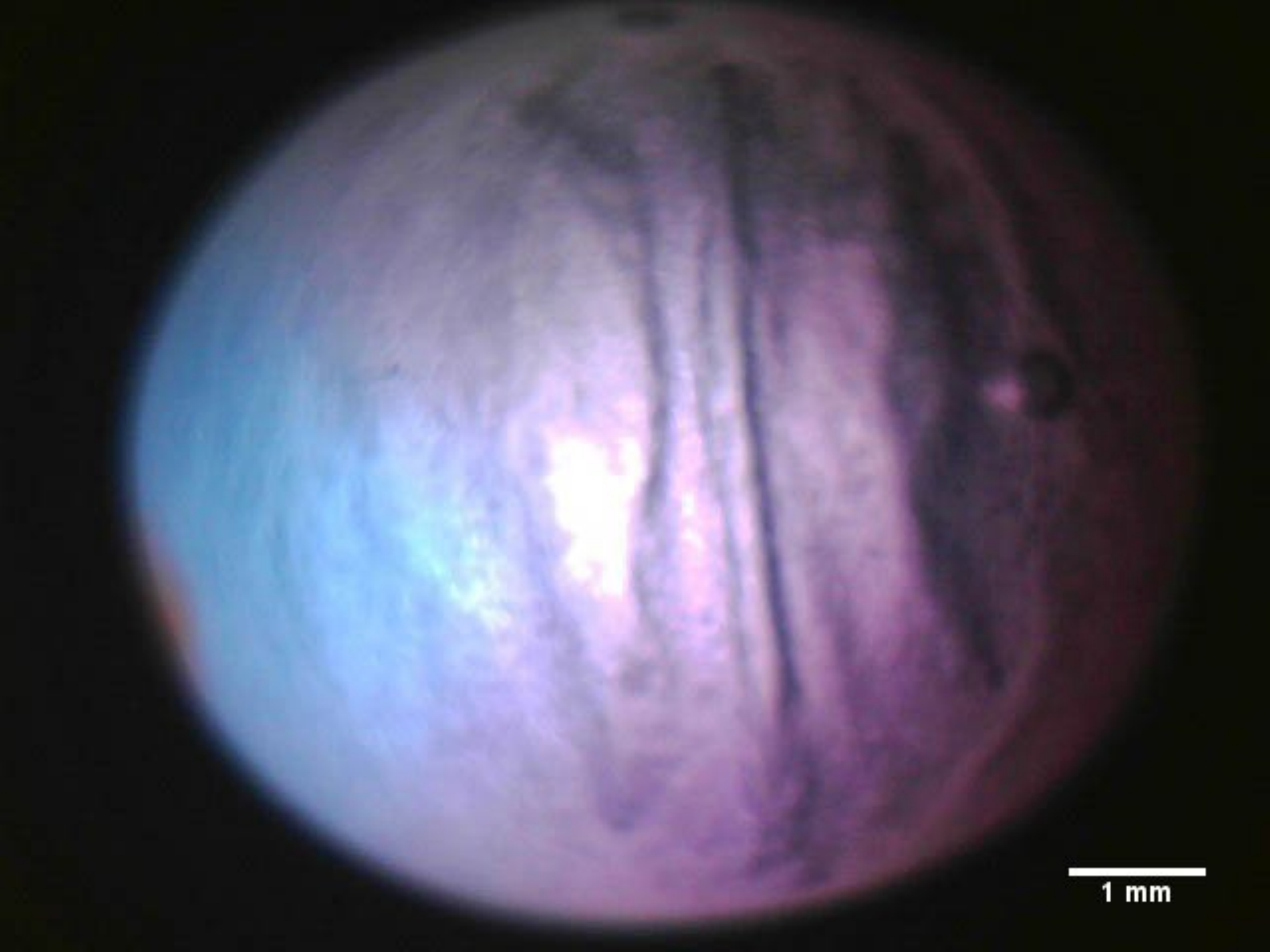}}
        \caption{A single \SI{50}{\micro\l} PTFE-coated BZ LM showing the LM at the start of the recording (a) and showing the 1st oxidation wave observed (b--f). The times images were taken are indicated in the captions for each sub-figure}
    \label{fig:PTFE50ul}
\end{figure}

Fig.~\ref{fig:PTFE50ul} shows the propagation of the 1st travelling wave observed in a single \SI{50}{\micro\litre} PTFE-coated BZ LM. For the wave to propagate across the 1st and 2nd half of the LM, it took on average 12 $\pm$ 2~s and 17 $\pm$ 4~s respectively, with a full oscillation taking on average 29 $\pm$ 5~s. 10 travelling waves were visible. Some centralised oscillations occurred at the start of the experiment, shown in Fig.~\ref{fig:PTFE50ul}b. Travelling waves did not appear until ca. 11~mins after the start of the experiment, propagating from right to left across the LM, shown in Fig.~\ref{fig:PTFE50ul}b--f. The LM started buckling after only ca. 2~mins after marble positioning. In Fig.~\ref{fig:PTFE50ul}e it appears the travelling wave splits into two wave-fronts, potentially arising from the buckling of the LM coating. Some gas evolution can be observed, shown by the trapped bubble in Fig.~\ref{fig:PTFE50ul}b--f. Full oxidation of the ferroin to ferriin had occurred within the marble after ca. 25 mins.

\begin{figure}[!tbp]
\centering
    \subfigure[840s]{\includegraphics[width=0.3\textwidth]{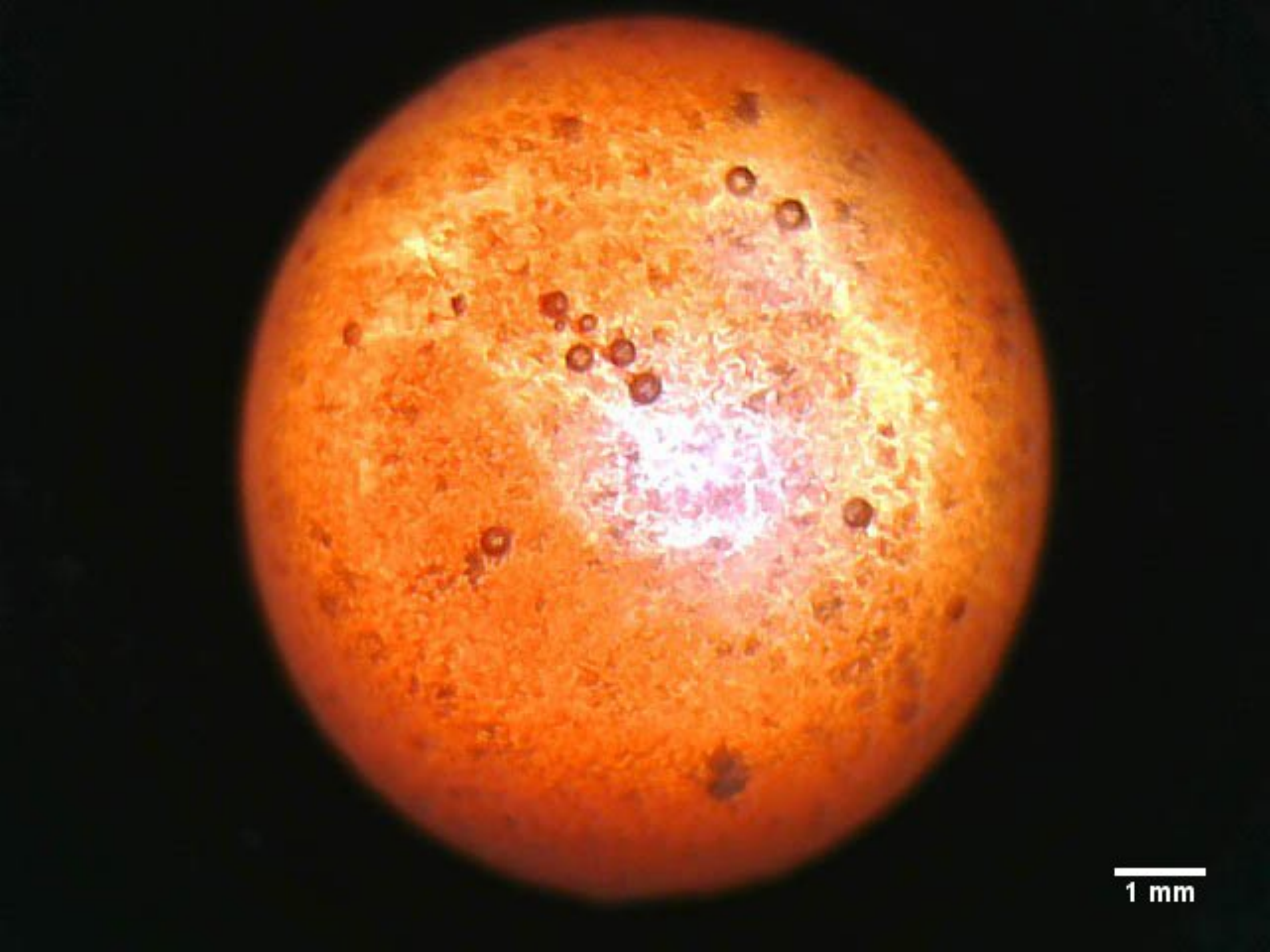}}
    \subfigure[850s]{\includegraphics[width=0.3\textwidth]{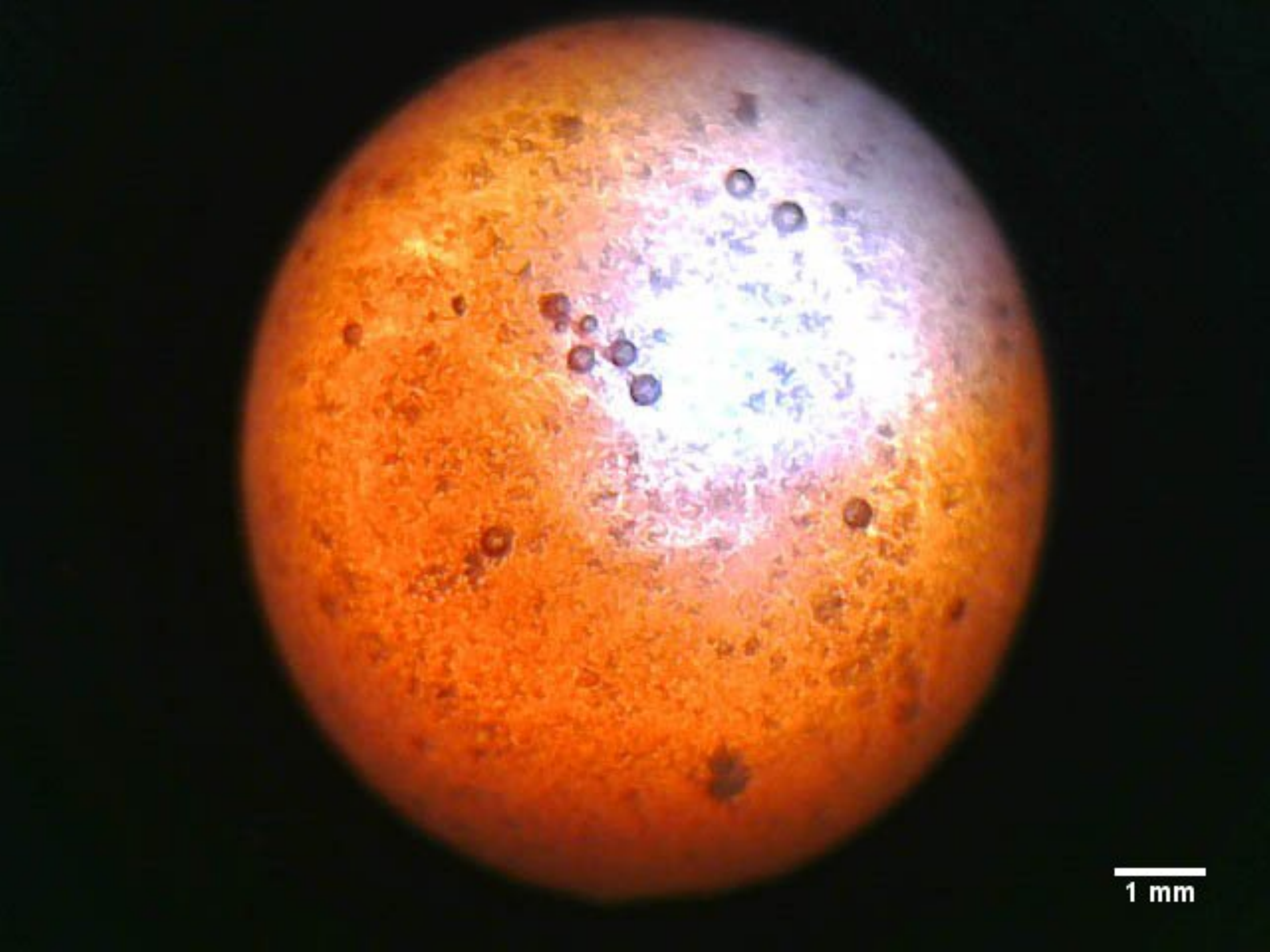}}
    \subfigure[860s]{\includegraphics[width=0.3\textwidth]{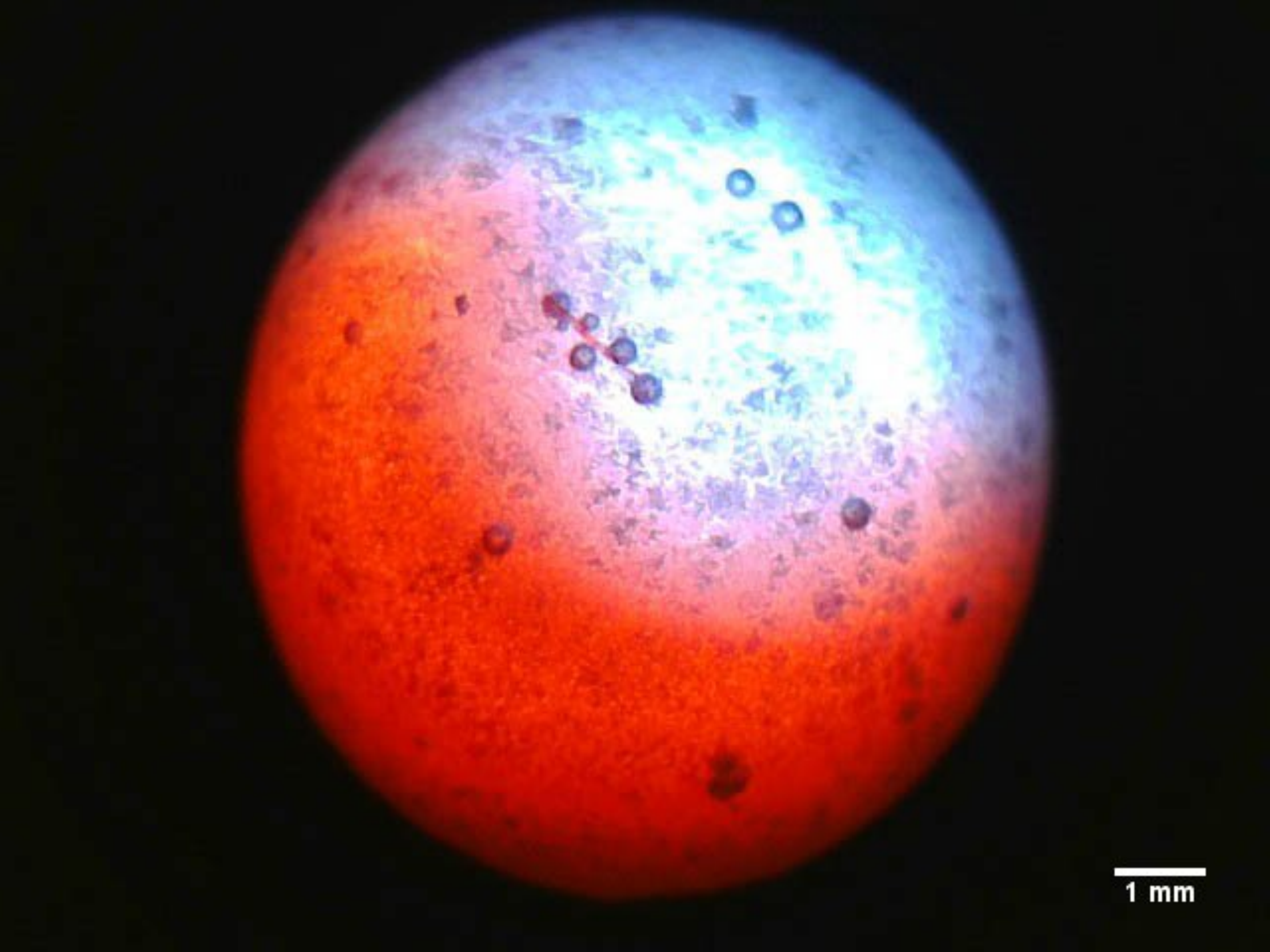}}
    \subfigure[865s]{\includegraphics[width=0.3\textwidth]{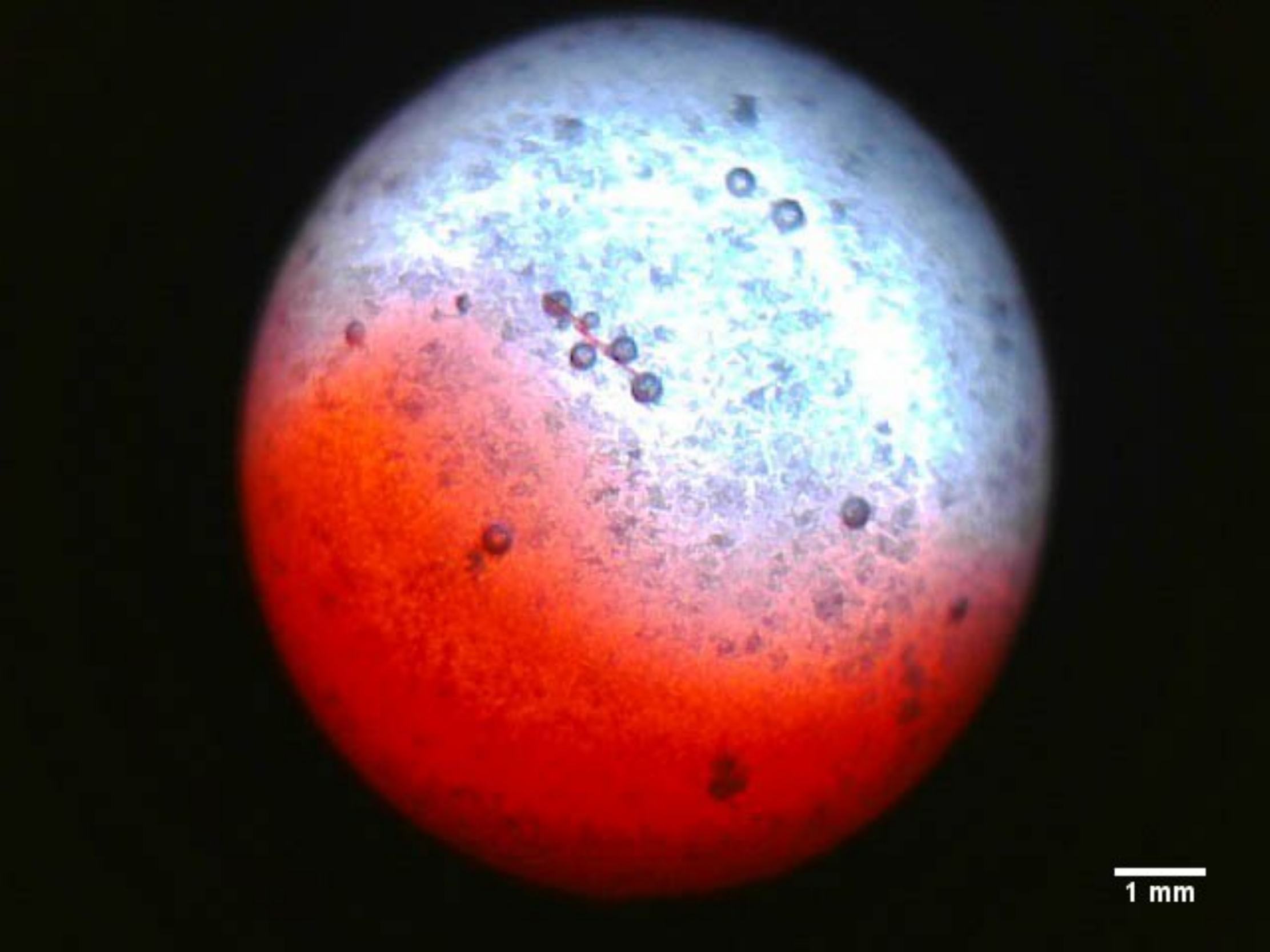}}
    \subfigure[870s]{\includegraphics[width=0.3\textwidth]{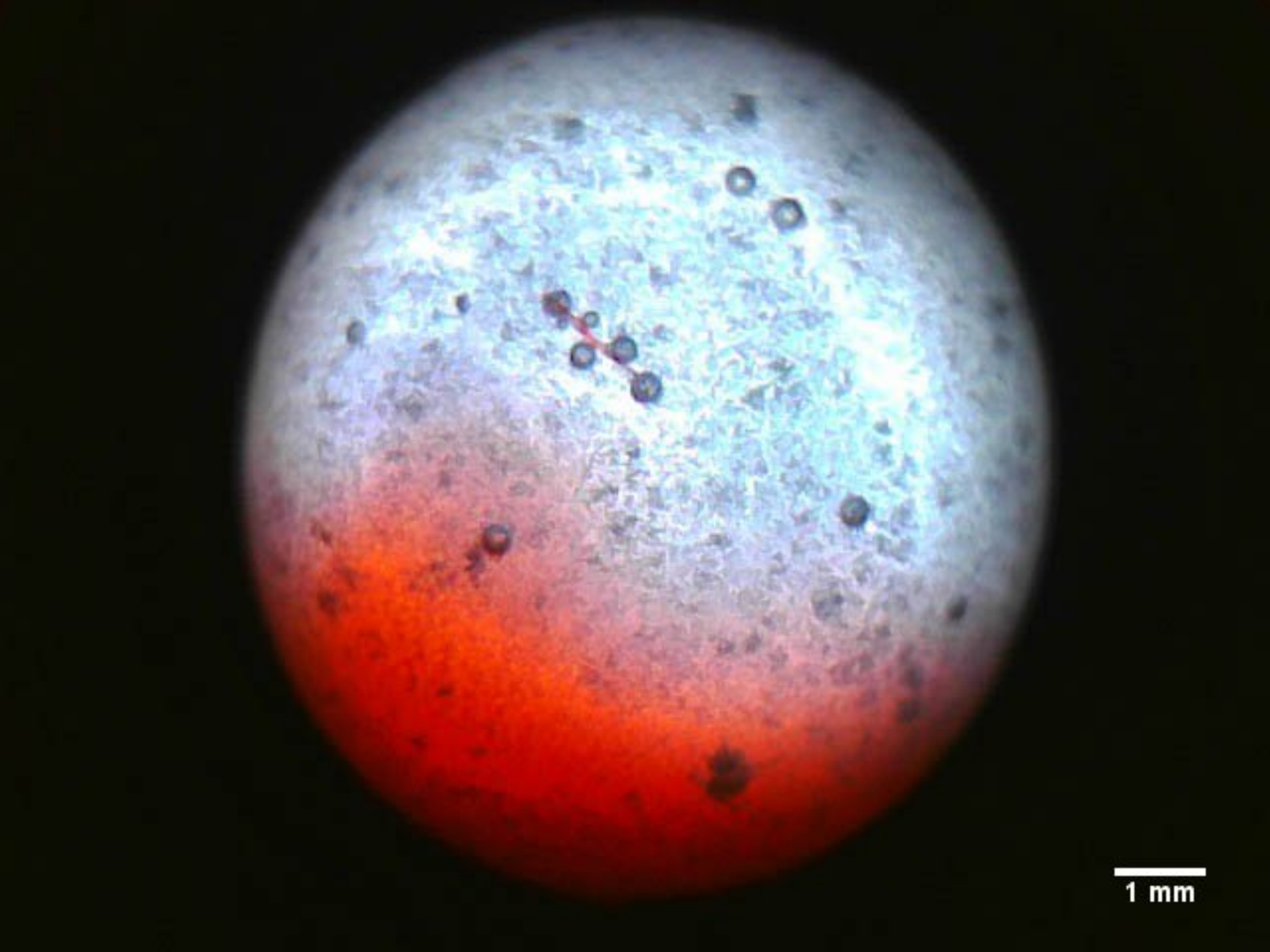}}
    \subfigure[875s]{\includegraphics[width=0.3\textwidth]{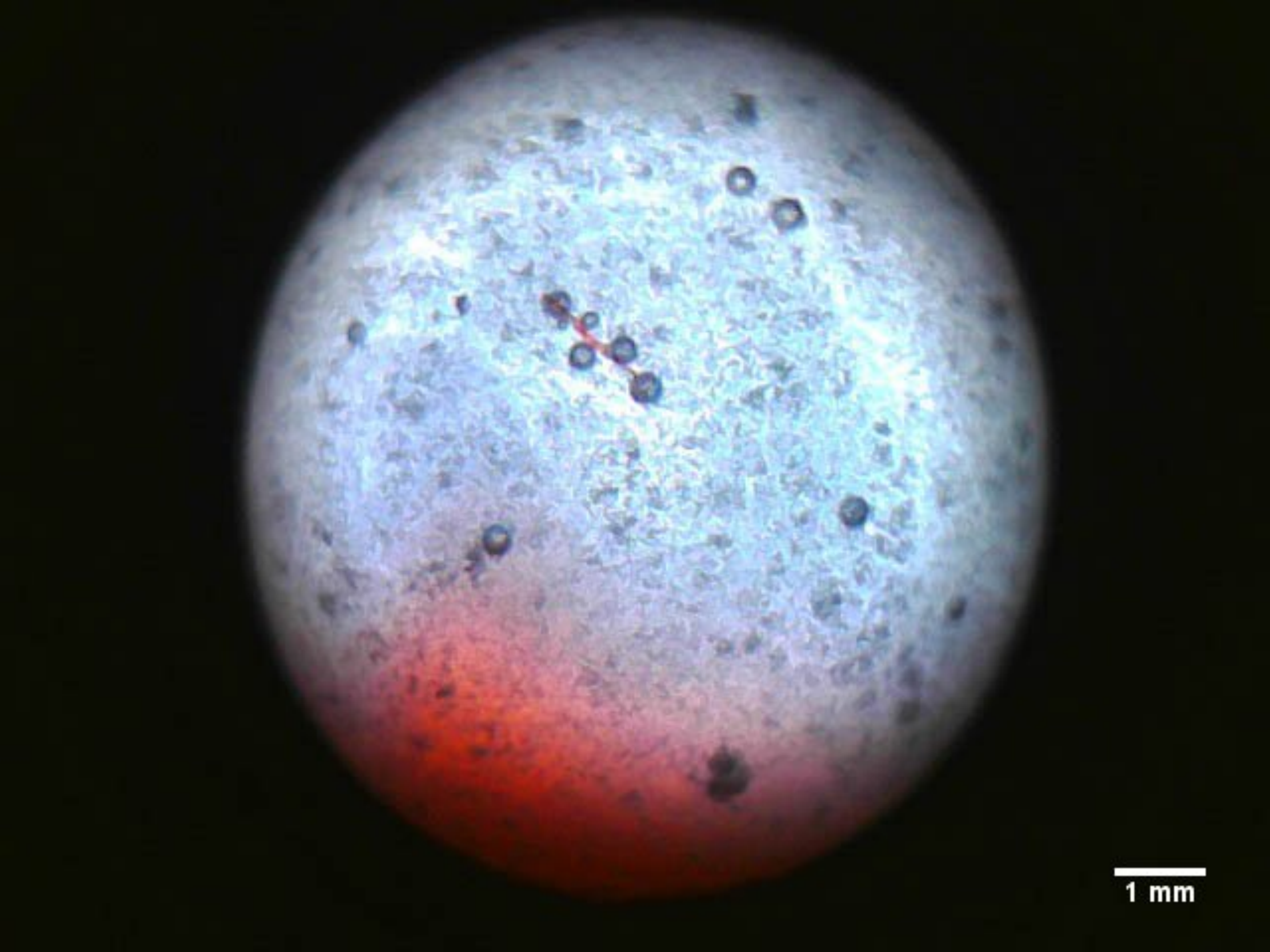}}
        \caption{A single \SI{100}{\micro\l} PTFE-coated BZ LM showing the LM at the start of the recording (a) and showing the only wave observed (b--f). The times images were taken are indicated in the captions for each sub-figure}
    \label{fig:PTFE100ul}
\end{figure}

Fig.~\ref{fig:PTFE100ul} shows the propagation of the only travelling wave observed in a single \SI{100}{\micro\litre} PTFE-coated BZ LM. For the wave to propagate across the 1st and 2nd half of the LM, it took 19~s and 21~s  respectively, with the full oscillation taking 40~s. Only one oscillation was observed, due to taking significantly longer to prepare a viable PTFE-coated BZ LM to record. Therefore, due to this and the fast onset of buckling, no further experiments were performed on PTFE-coated BZ LMs.

\begin{figure}[!tbp]
\centering
    \subfigure[312s]{\includegraphics[width=0.3\textwidth]{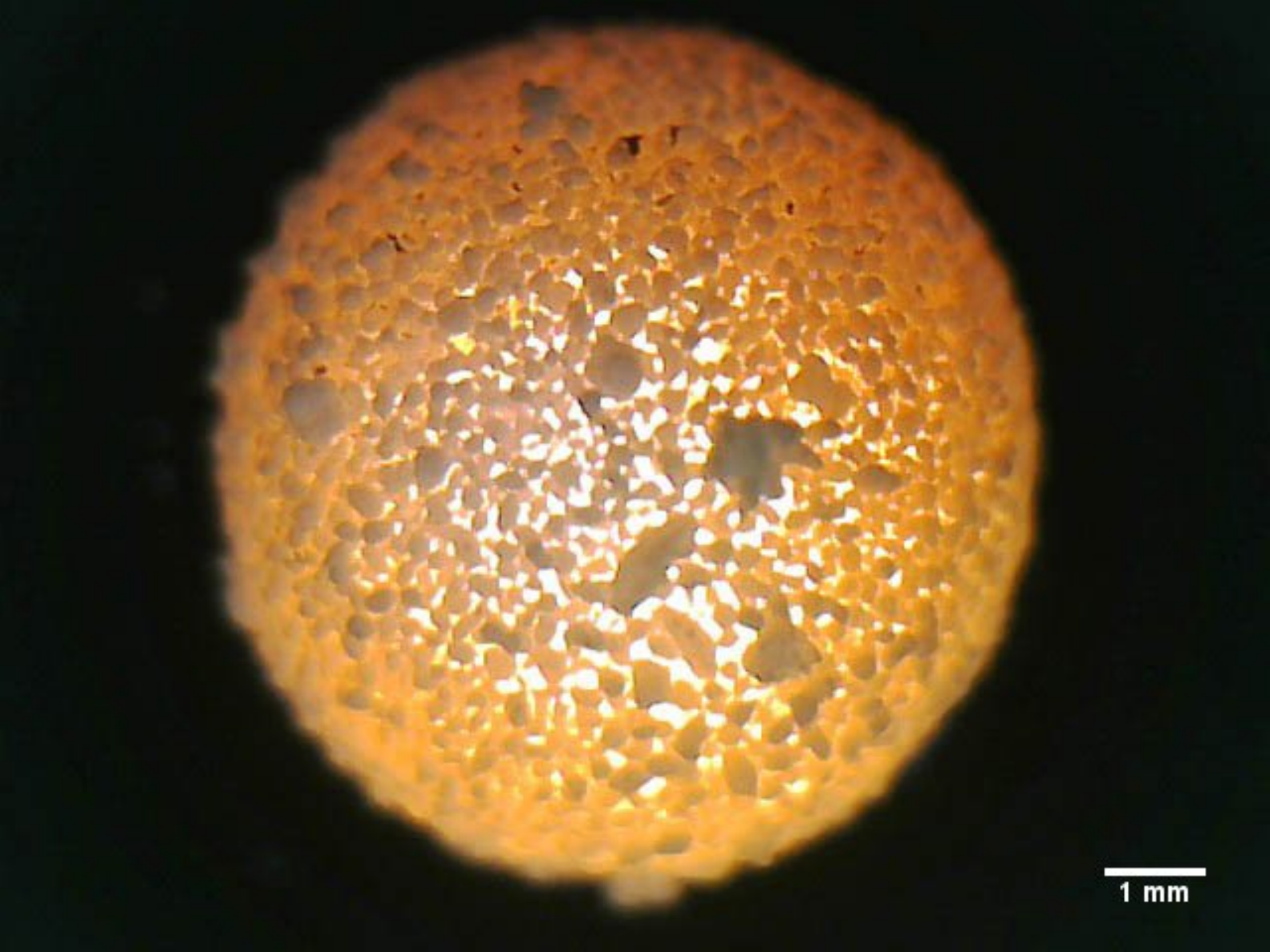}}
    \subfigure[324s]{\includegraphics[width=0.3\textwidth]{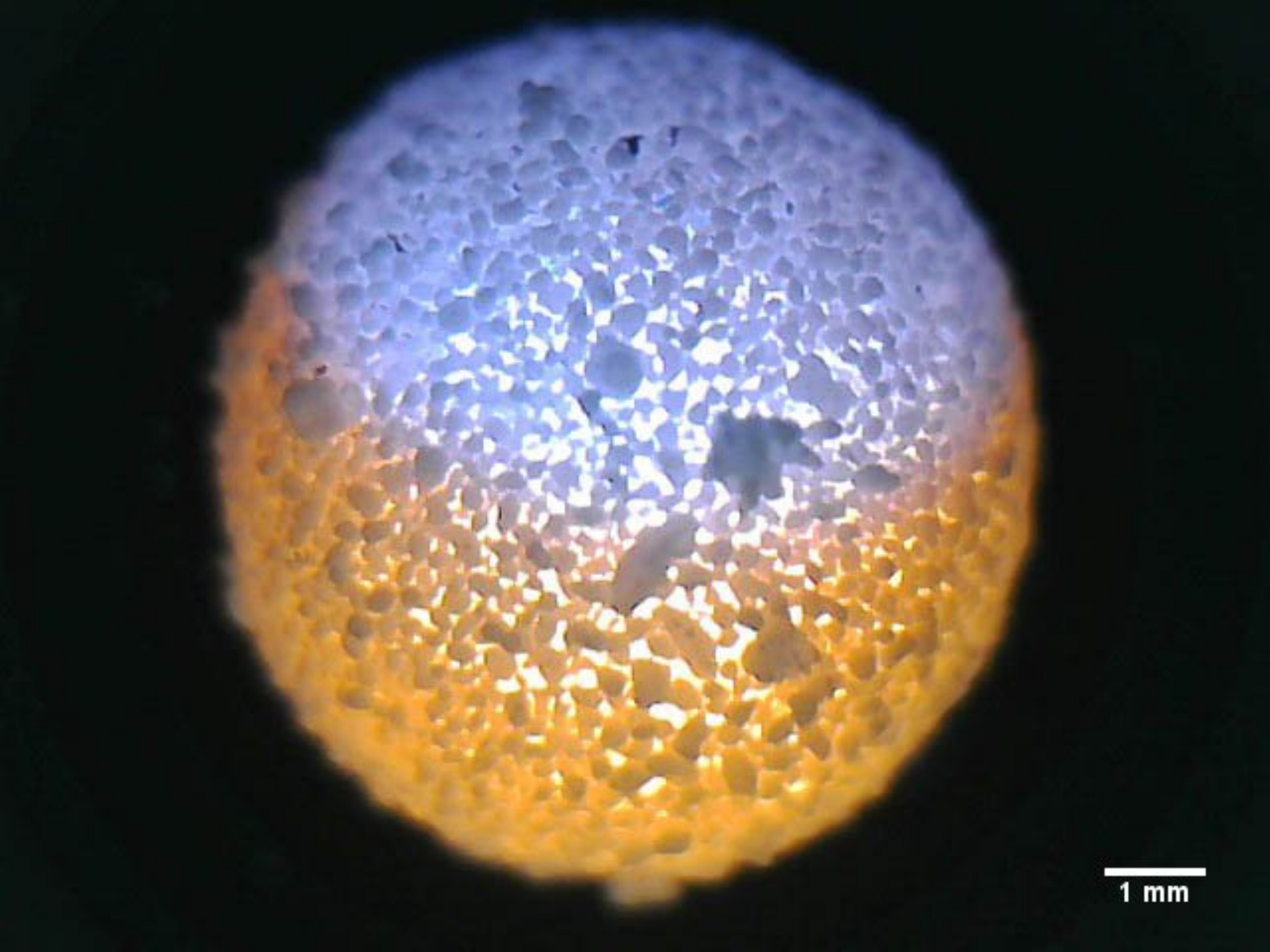}}
    \subfigure[337s]{\includegraphics[width=0.3\textwidth]{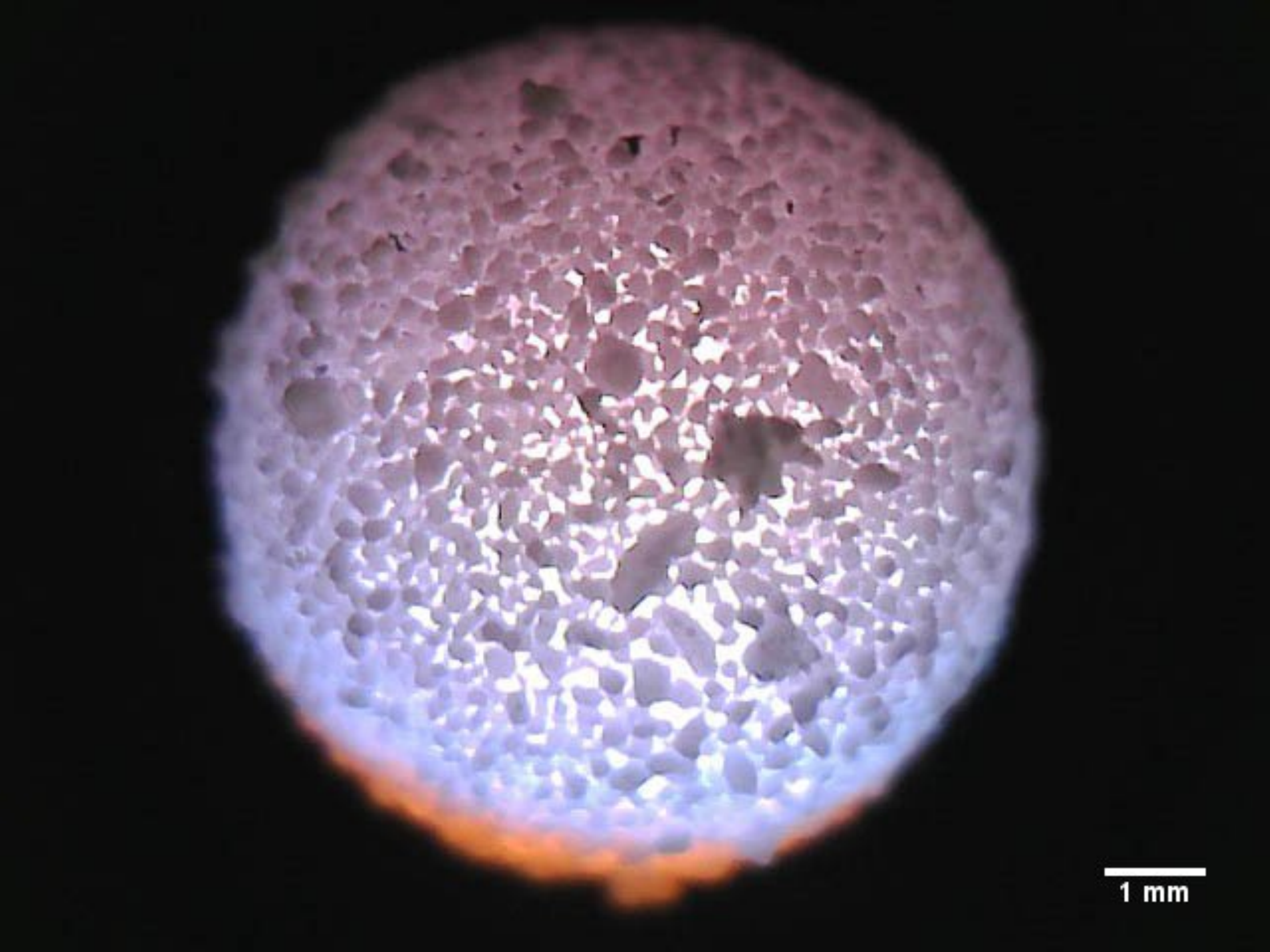}}
    \subfigure[526s]{\includegraphics[width=0.3\textwidth]{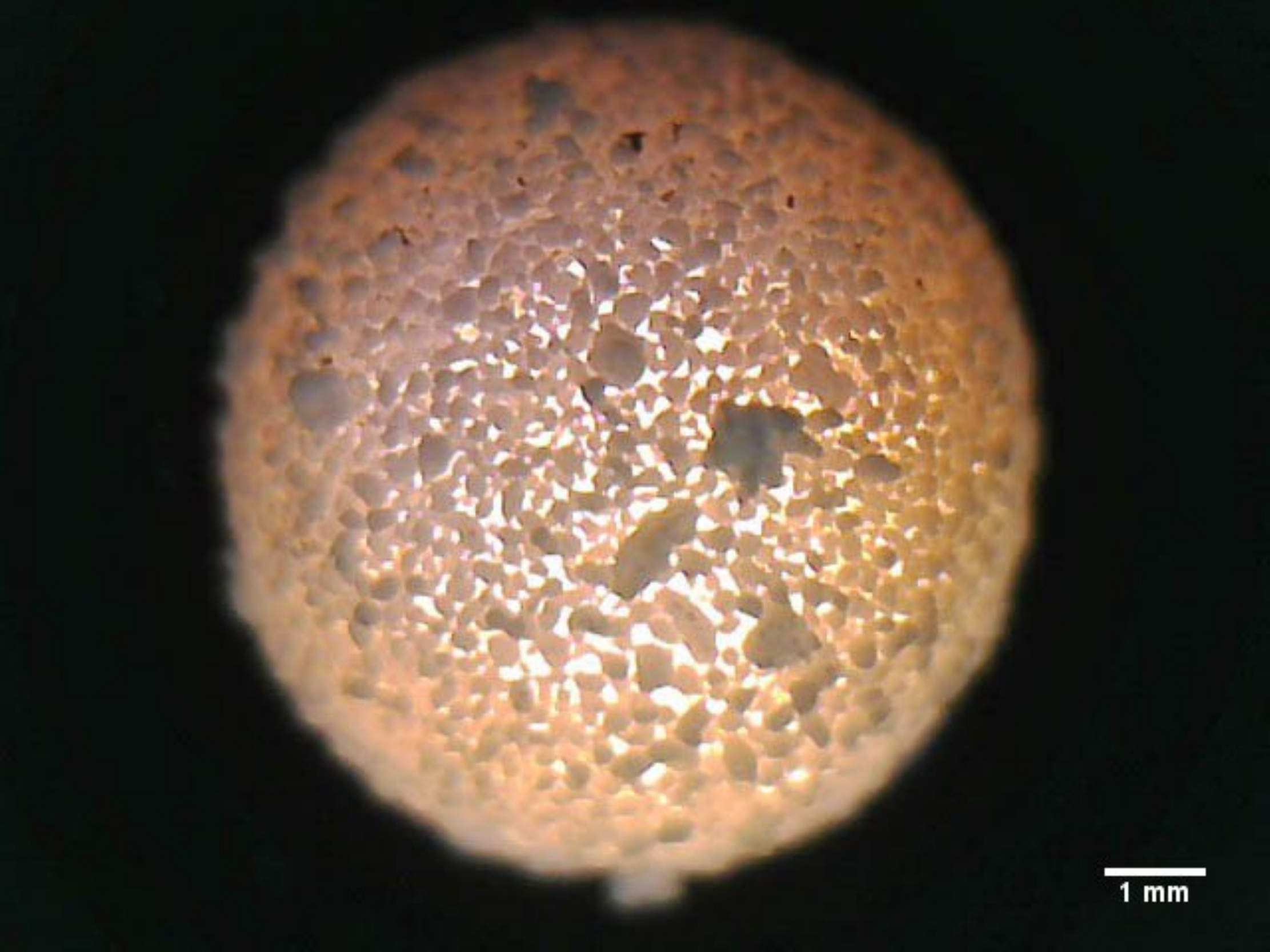}}
    \subfigure[538s]{\includegraphics[width=0.3\textwidth]{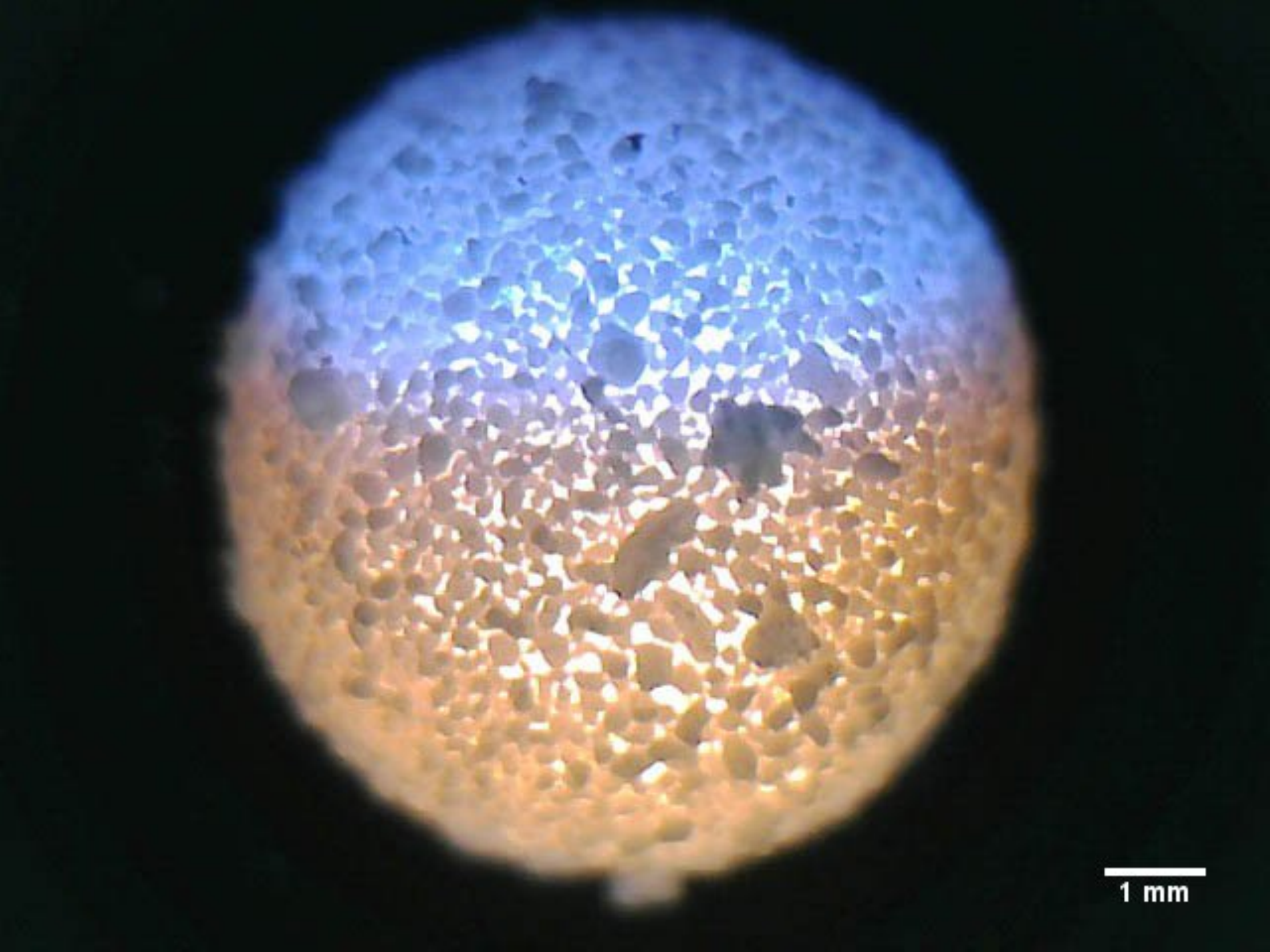}}
    \subfigure[552s]{\includegraphics[width=0.3\textwidth]{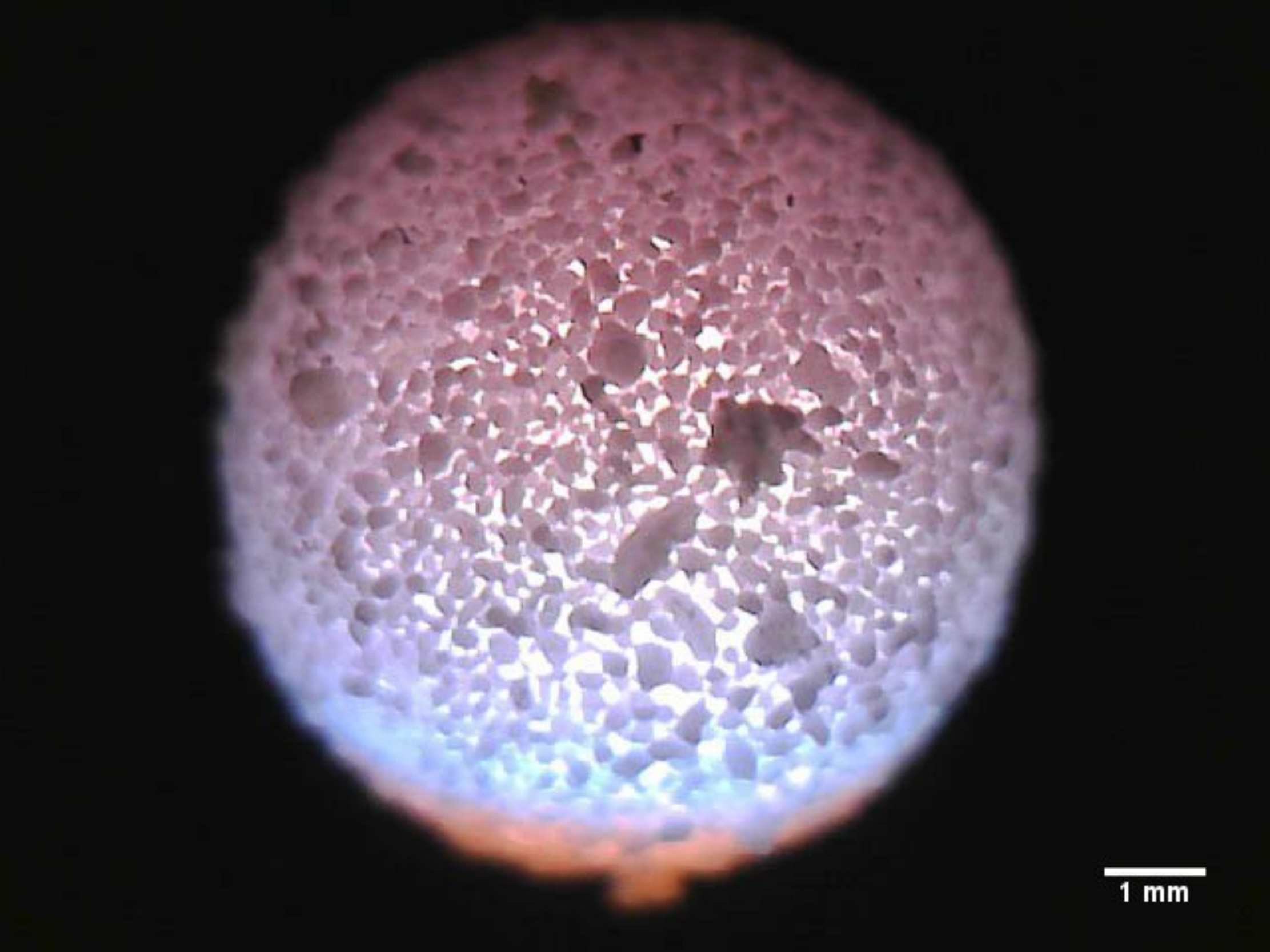}}
        \caption{A single \SI{50}{\micro\l} PE-coated BZ LM showing the 1st (a -- c) and 2nd (d -- f) oxidation waves. The times images were taken are indicated in the captions for each sub-figure}
    \label{fig:PE50ul}
\end{figure}

Fig.~\ref{fig:PE50ul} shows the 1st and 2nd travelling waves observed in a single \SI{50}{\micro\litre} PE-coated BZ LM. For the wave to propagate across the 1st and 2nd half of the LM, it took on average 10 $\pm$ 2~s for both, with a full oscillation taking on average 20 $\pm$ 3~s to travel across the diameter of the LM. 55 single travelling waves were visible, which moved across the LM from top to bottom. After ca. 2~mins small movements of the LM was observed. However, it was hard to judge whether this was a result of variations in inter-facial tension due to the travelling waves, previously observed in BZ droplets in oil~\cite{Steinbock1998, Kitahata2002, Kitahata2012} or simply the movement of the powder coating due to evaporation of the encapsulated BZ solution.  After ca. 18~mins, buckling of the \SI{50}{\micro\litre} PE-coated BZ LM was observed. After ca. 34~mins some gas evolution \ce{CO2} and /or \ce{CO} was observed under the coating, however the bubbles did not affect the travelling waves as most had been observed by this time. Multiple oscillations occurred after ca. 36~mins, at which time significant buckling of the marble had occurred. Full oxidation of the ferroin to ferriin had occurred within the marble after ca. 55~mins.

In a repeat experiment, for the wave to propagate across the 1st and 2nd half of the LM, it took on average 14 $\pm$ 1~s and 12 $\pm$ 2~s respectively, with a full oscillation taking on average 25 $\pm$ 2~s. 32 single travelling waves were visible, slightly less than the previous \SI{50}{\micro\litre} LM, attributed to the preparation and setup time of the LM underneath the camera. The travelling waves in this LM were observed to move across the LM from right to left. After ca. 19~mins, buckling was observed, the same as the previous \SI{50}{\micro\litre} LM. After ca. 39~mins again some gas evolution occurred. After ca. 45~mins the travelling waves were not easy to distinguish and multiple oscillations started occurring. Full oxidation of the ferroin to ferriin had occurred within the LM after ca. 49~mins. After this length of time significant size gas bubbles can be observed trapped under the buckled coating of the LM.

\begin{figure}[!tbp]
\centering
    \subfigure[597s]{\includegraphics[width=0.3\textwidth]{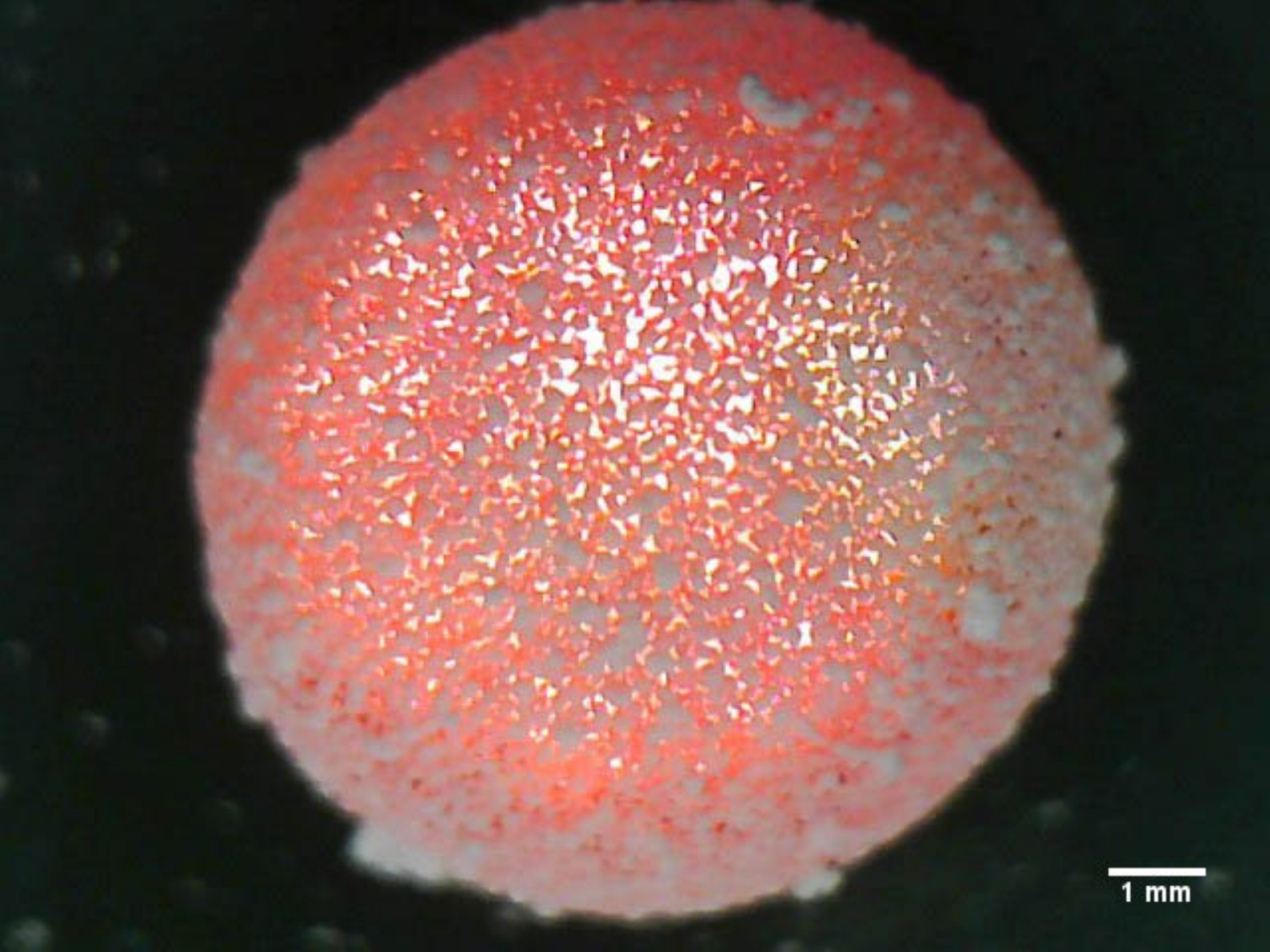}}
    \subfigure[609s]{\includegraphics[width=0.3\textwidth]{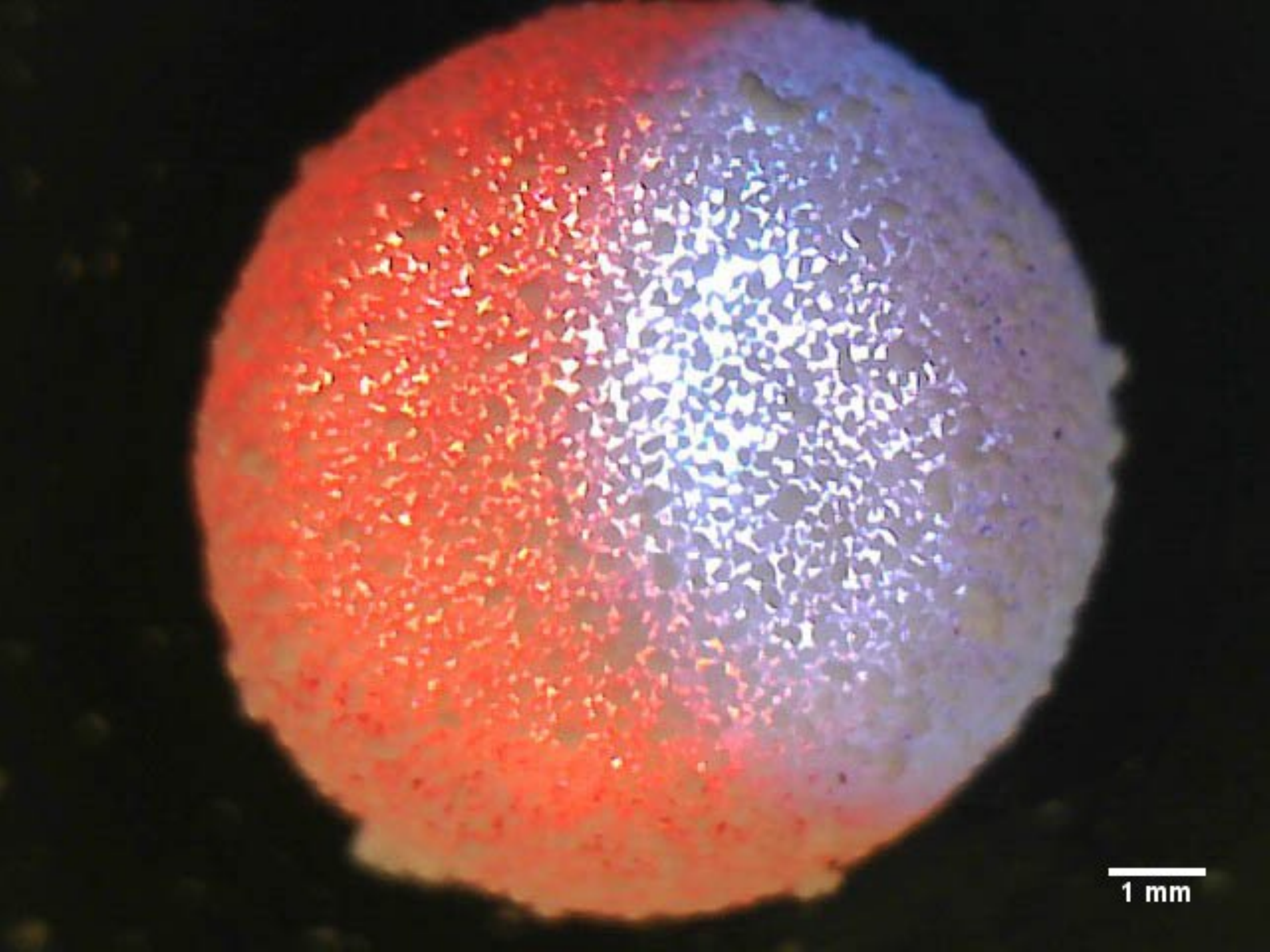}}
    \subfigure[632s]{\includegraphics[width=0.3\textwidth]{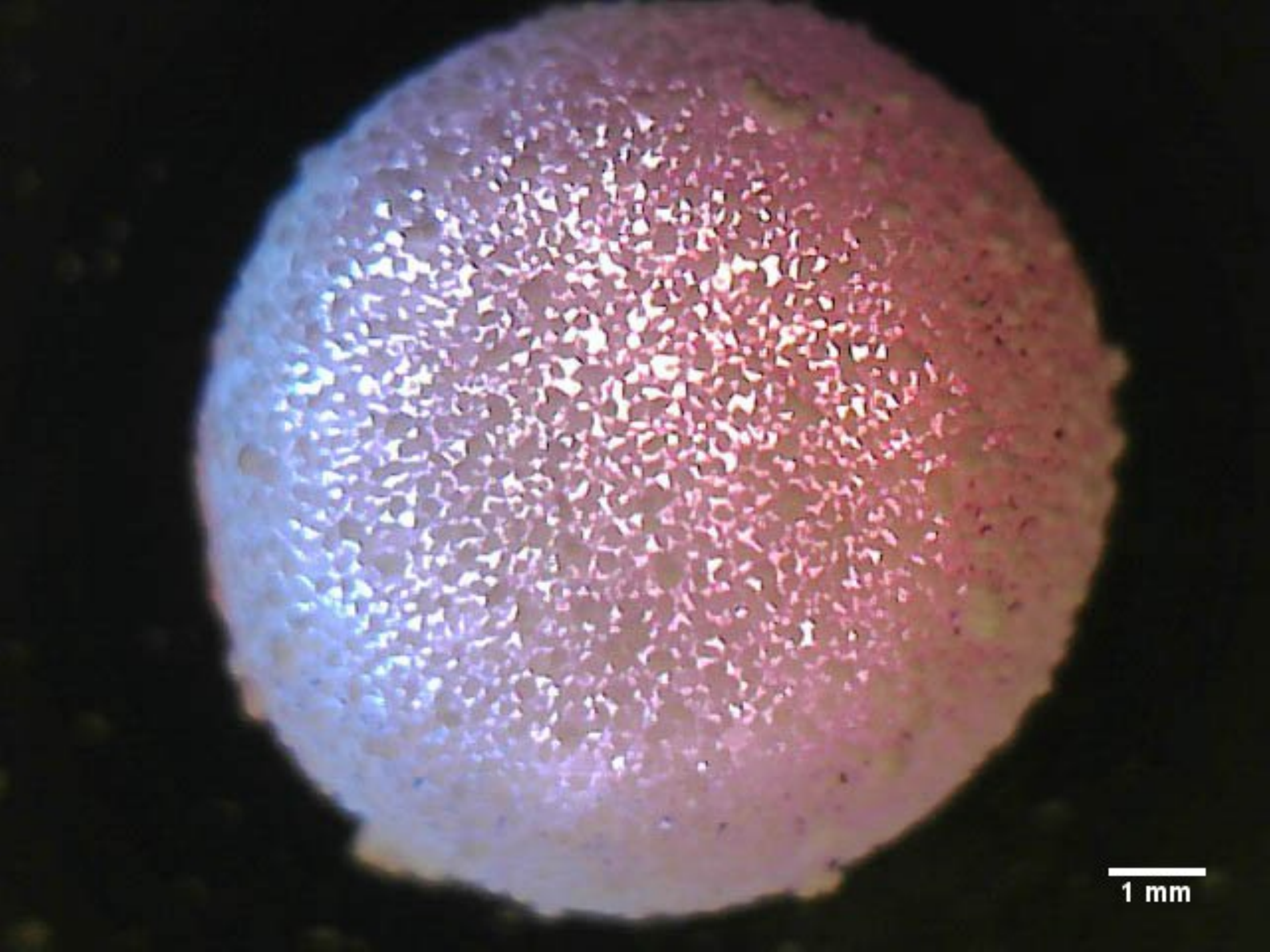}}
    \subfigure[1268s]{\includegraphics[width=0.3\textwidth]{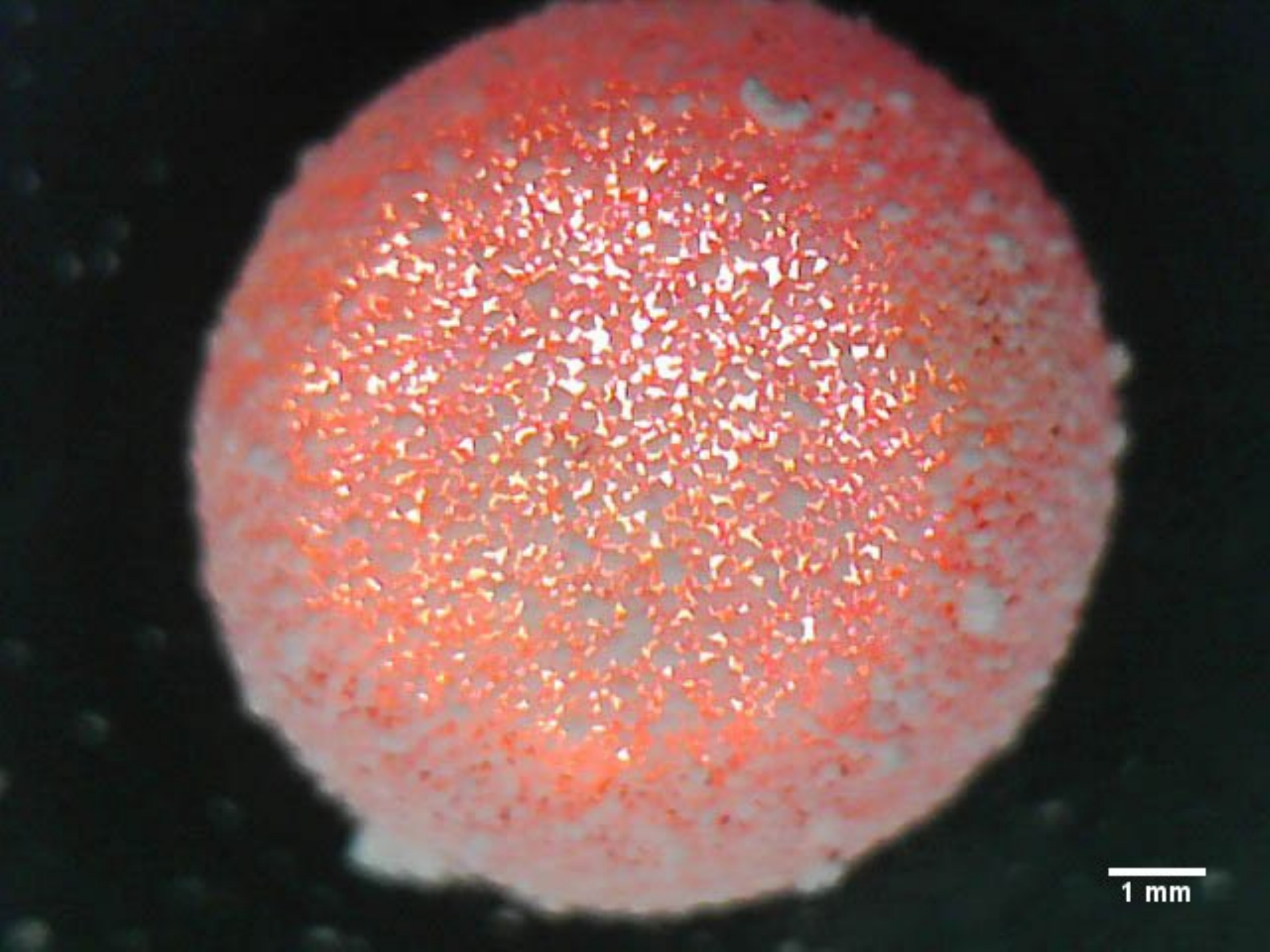}}
    \subfigure[1286s]{\includegraphics[width=0.3\textwidth]{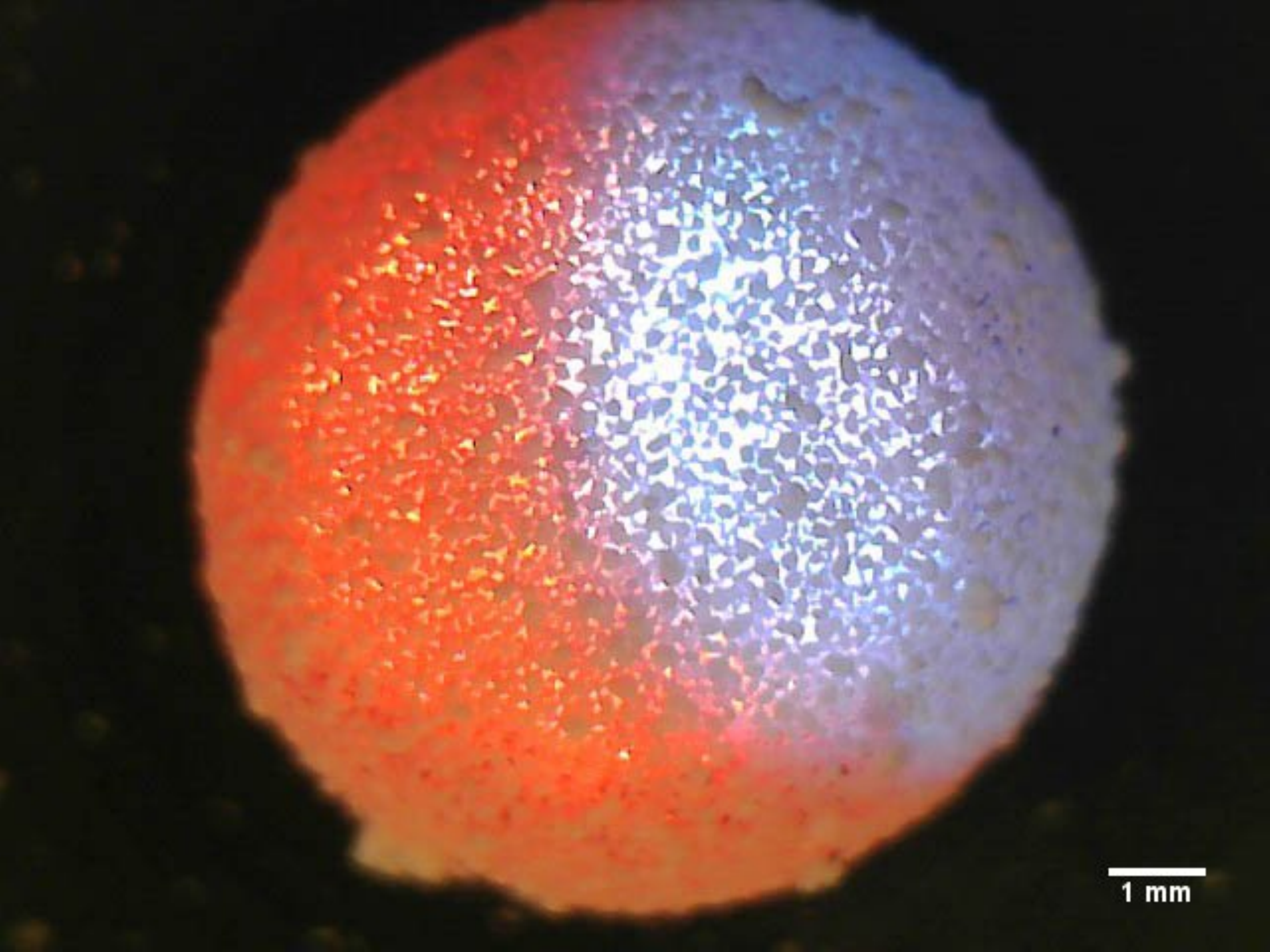}}
    \subfigure[1309s]{\includegraphics[width=0.3\textwidth]{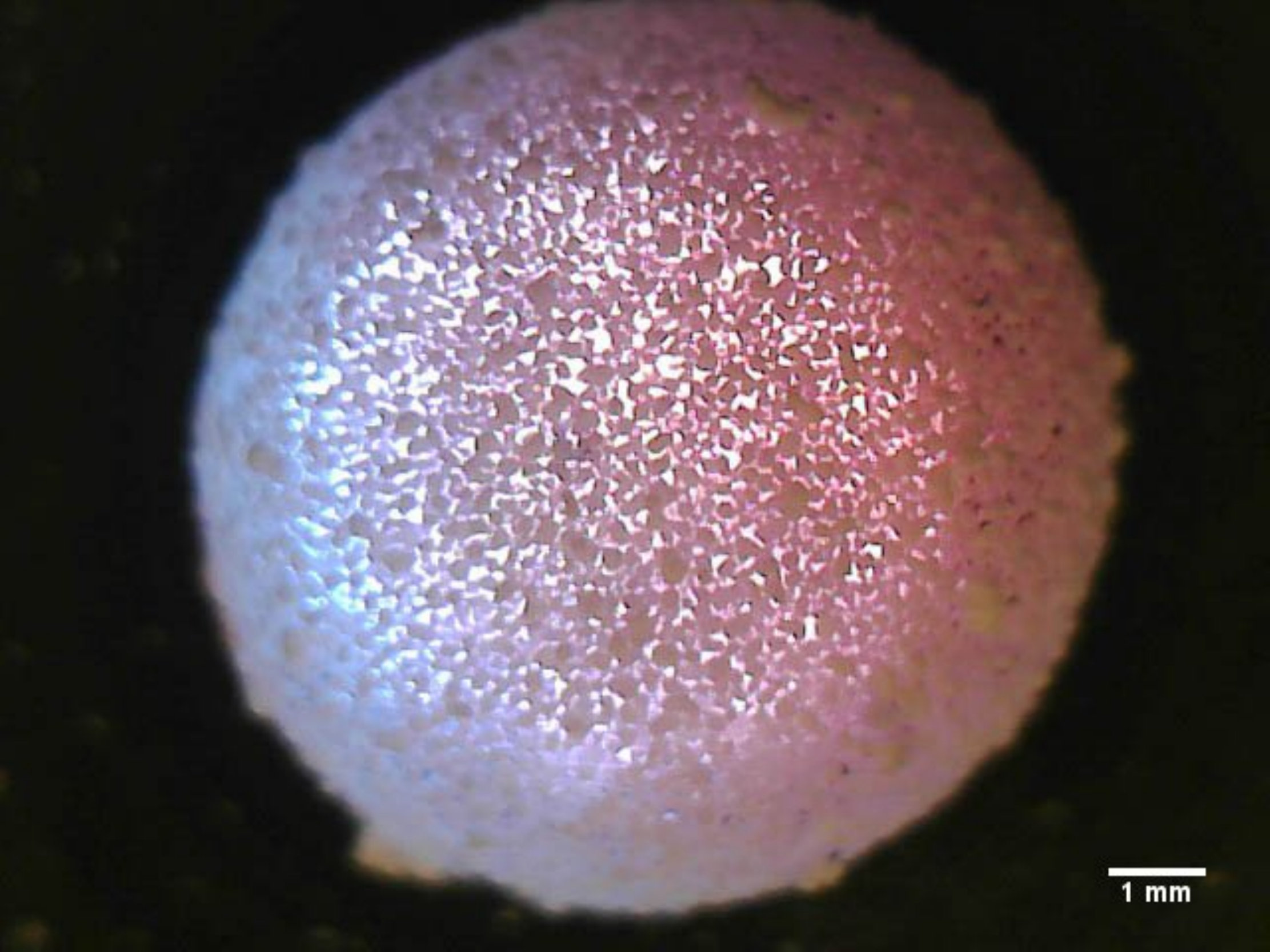}}
    \subfigure[1900s]{\includegraphics[width=0.3\textwidth]{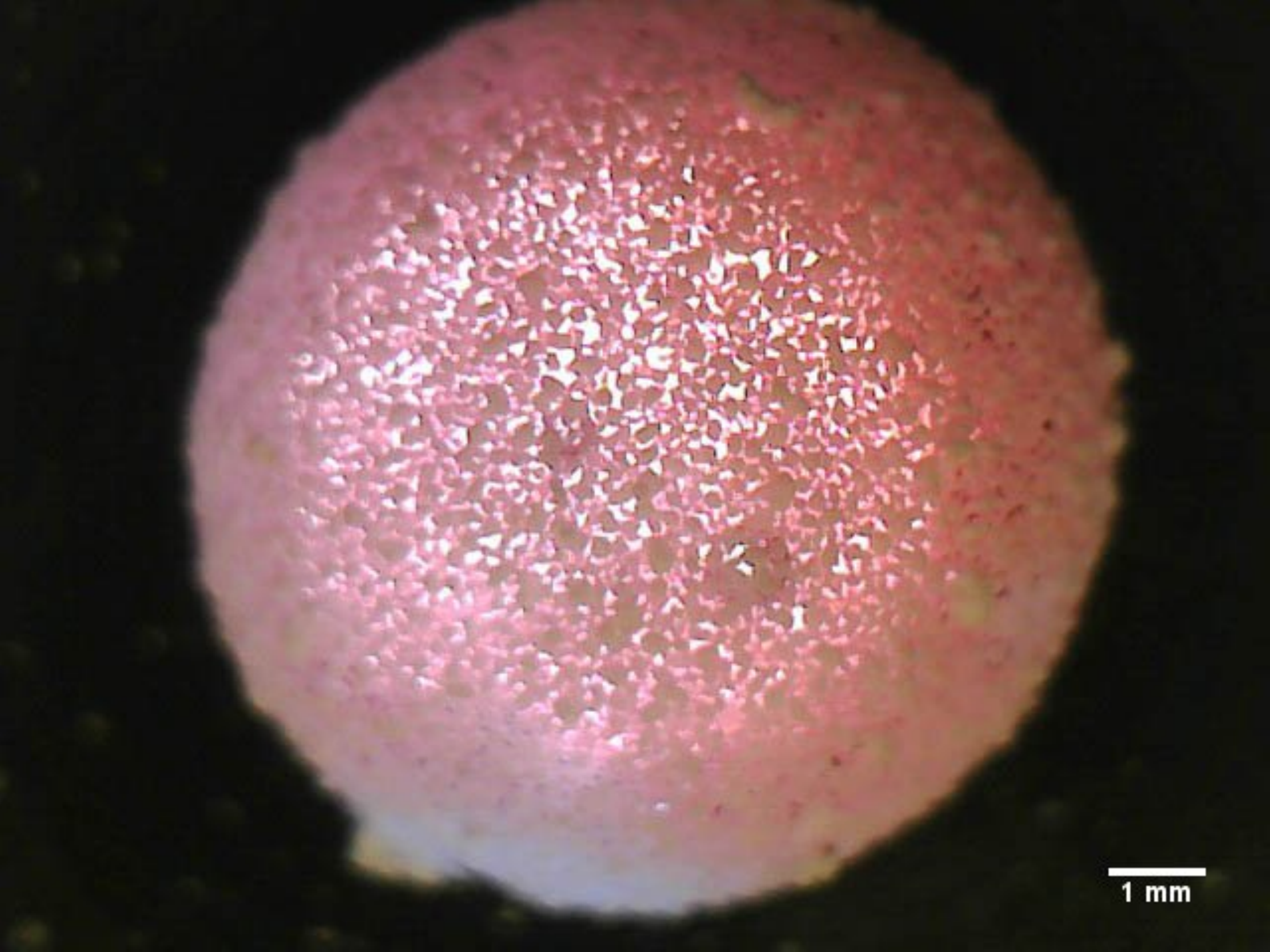}}
    \subfigure[1911s]{\includegraphics[width=0.3\textwidth]{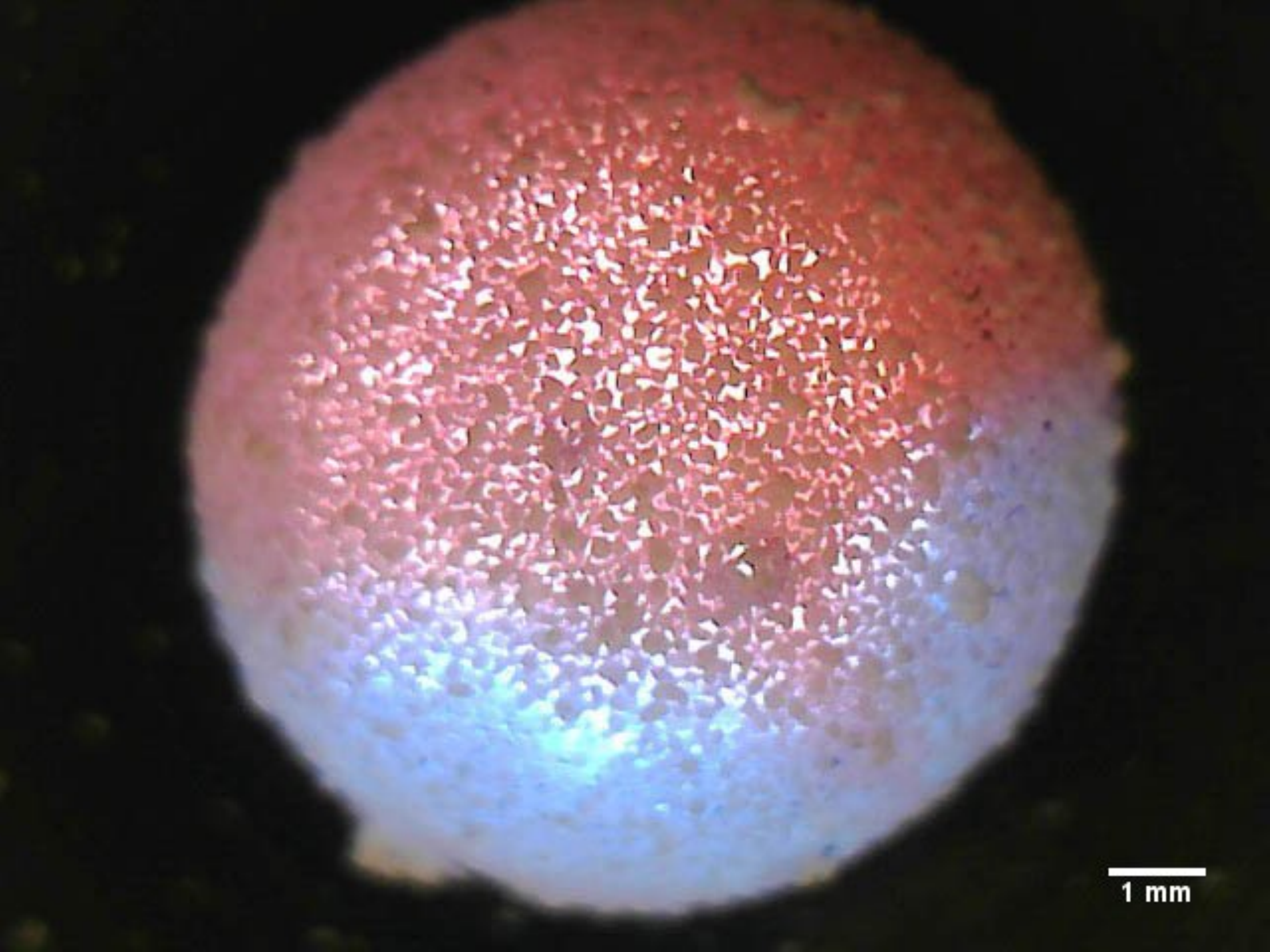}}
    \subfigure[1920s]{\includegraphics[width=0.3\textwidth]{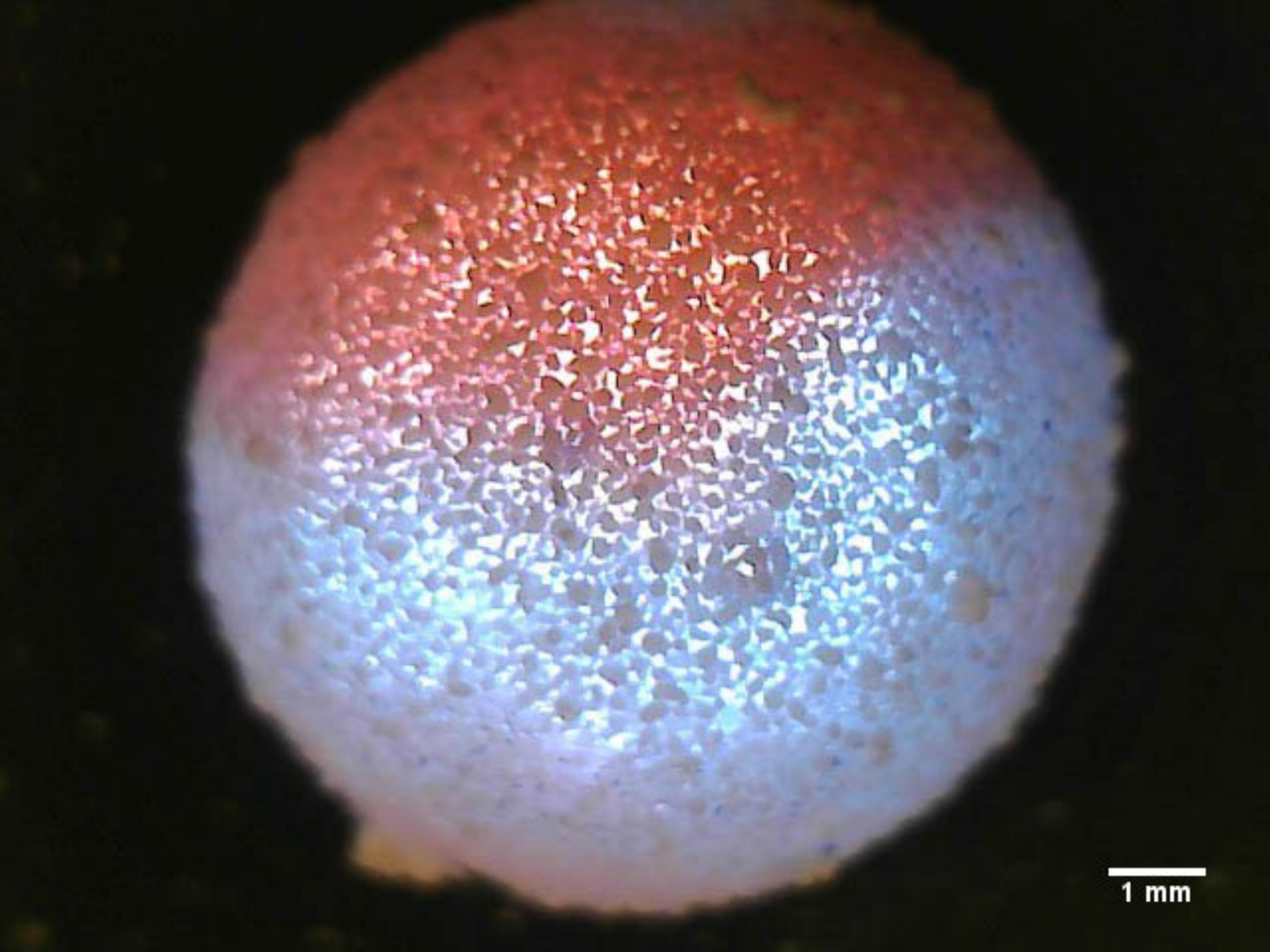}}
    \subfigure[1922s]{\includegraphics[width=0.3\textwidth]{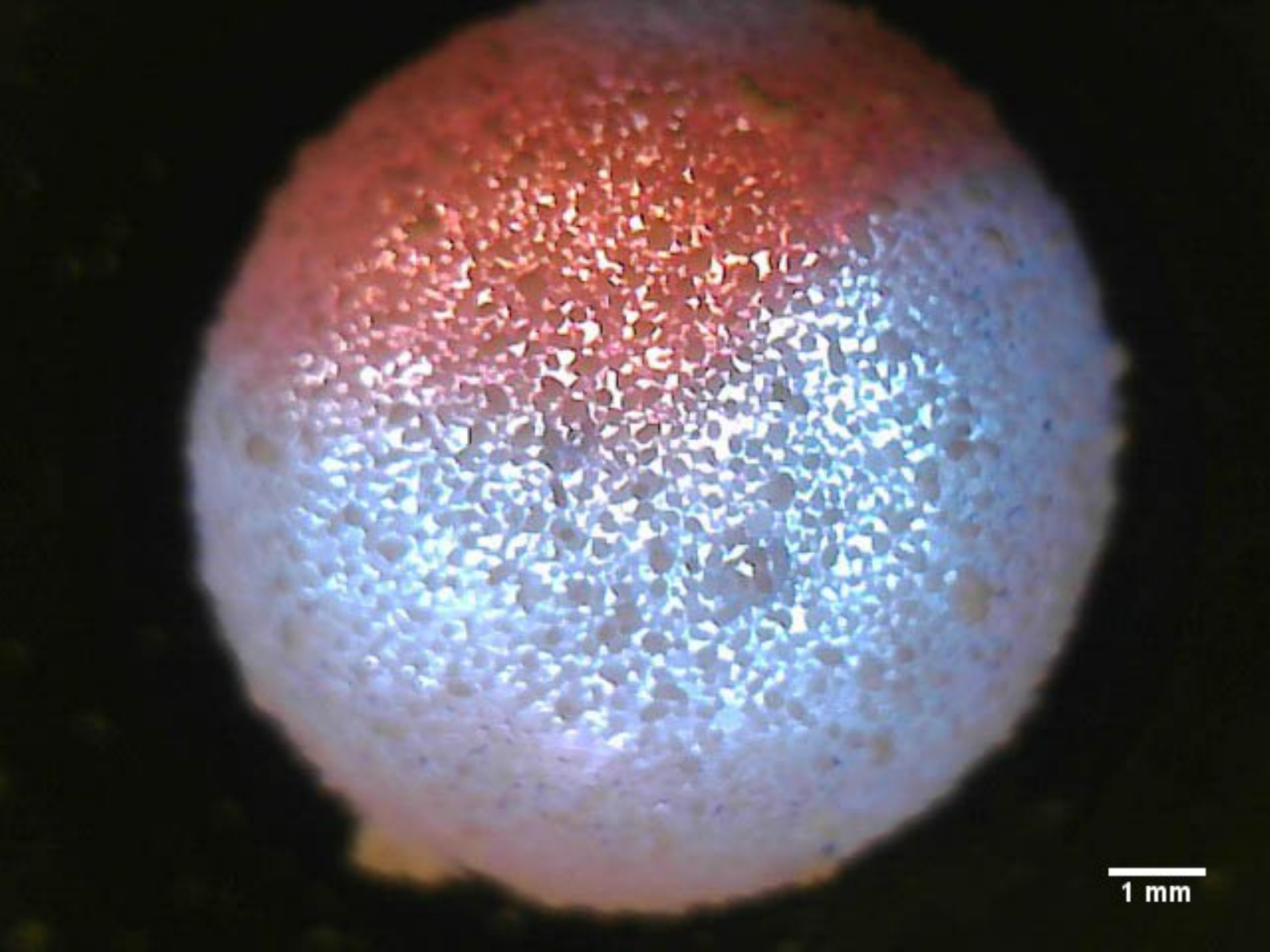}}
    \subfigure[1925s]{\includegraphics[width=0.3\textwidth]{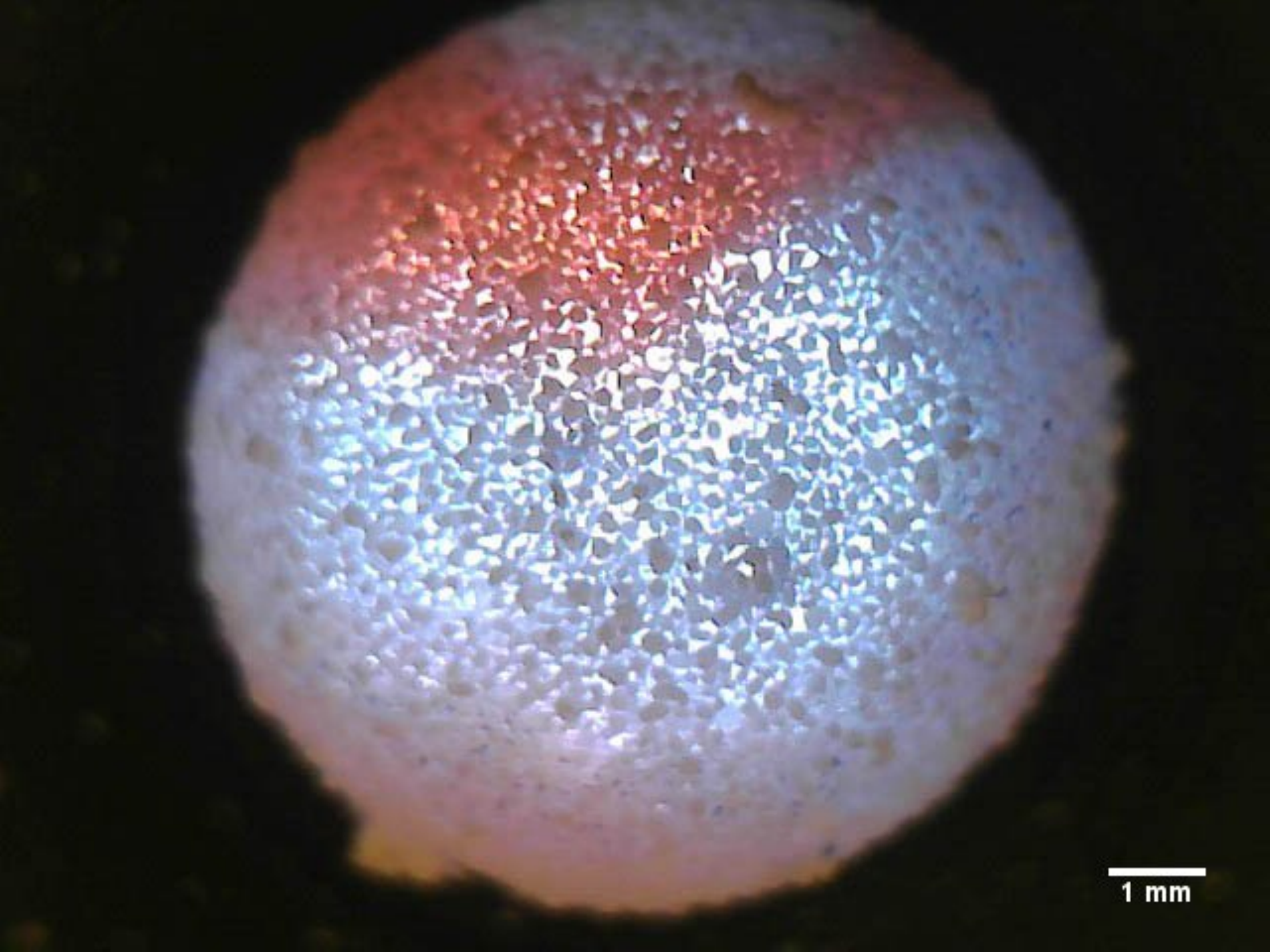}}
    \subfigure[1933s]{\includegraphics[width=0.3\textwidth]{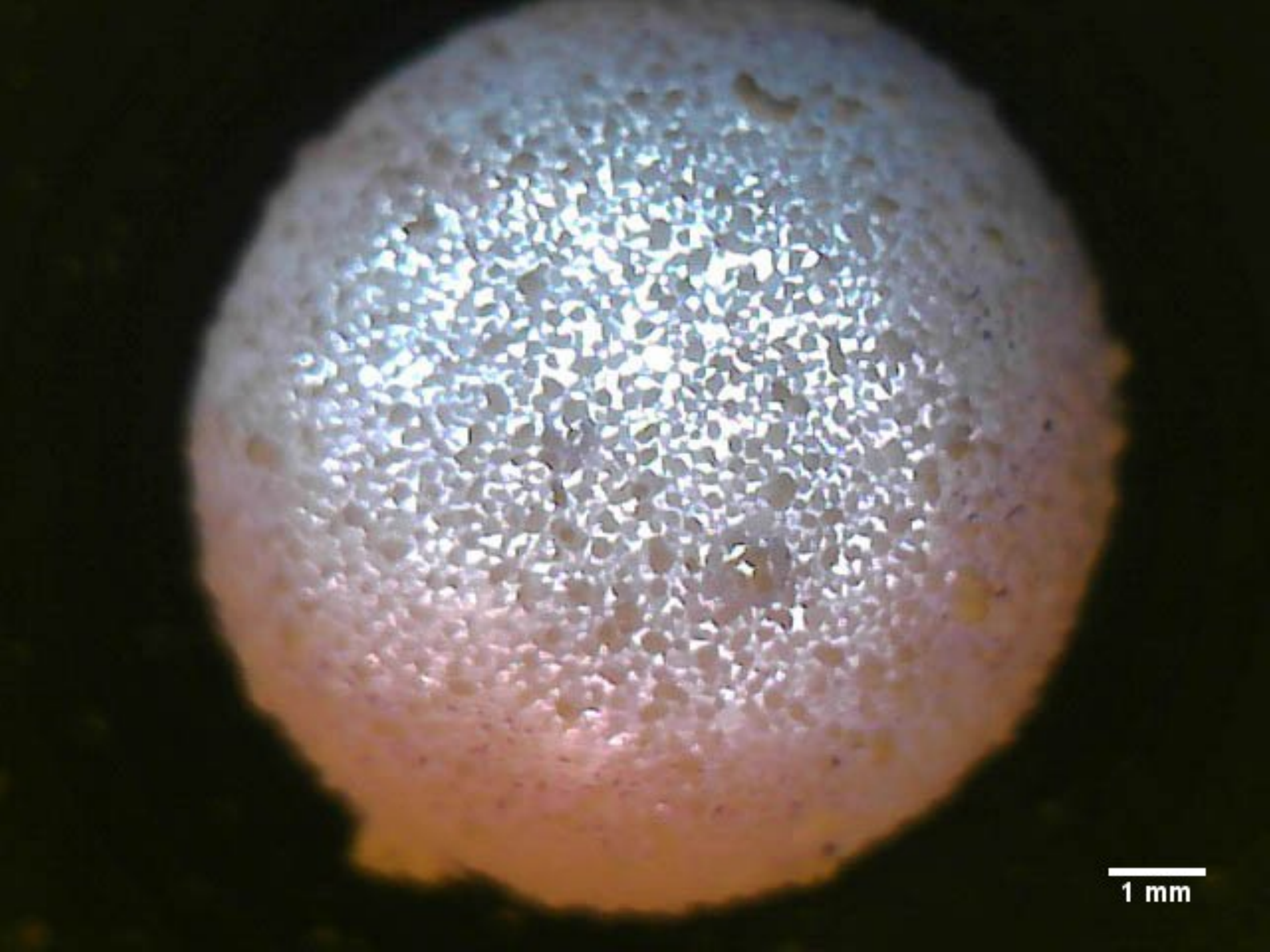}}
    \caption{A single \SI{100}{\micro\l} PE-coated BZ LM coated showing every oxidation wave observed. The times images were taken are indicated in the captions for each sub-figure}
    \label{fig:PE100ul}
\end{figure}

Fig.~\ref{fig:PE100ul} shows the travelling waves in a single \SI{100}{\micro\litre} PE-coated BZ LM. For the wave to move across the 1st and 2nd half of the LM, it took on average 18 $\pm$ 6~s and 22 $\pm$ 1~s respectively, with a full oscillation taking on average 40 $\pm$ 6~s. 5 single travelling waves were visible (at 1897s, 3 waves were observed at one time, shown in Fig.~\ref{fig:PE100ul}j--l), travelling from right to left across the LM. Gas evolution occurred after ca. 25~mins, slightly earlier than observed for \SI{50}{\micro\litre} LMs. Significant buckling of the LM occurred after ca. 30~mins. Full oxidation of the ferroin to ferriin has occured after ca. 42~mins.

In a repeat experiment, for the wave to propagate across the 1st and 2nd half of the LM, it took on average 16 $\pm$ 5~s and 19 $\pm$ 4~s respectively, with a full oscillation taking on average 35 $\pm$ 7~s. 6 single travelling waves were visible, propagating from the bottom to the top of the LM. After ca. 19~mins buckling of the \SI{100}{\micro\litre} PE-coated BZ LM was observed. Gas evolution occurred after ca. 25~mins, the same as the previous \SI{100}{\micro\litre} single BZ LM analysed. Multiple oscillations occurred after ca. 21~mins. Full oxidation of the ferroin to ferriin had occurred within the marble after ca. 42~mins. The smaller volume single LMs exhibited more visible oscillations than the larger volume LMs. 

Small vibrations, impacts and collisions caused both PE BZ LMs and PTFE BZ LMs to coalesce or burst relatively easily. PE-caoted LMs were easier to roll into position over either the LED to record a single LM or position into disordered and ordered arrays. The single BZ LM experiments proved which out of the two coatings selected would be a viable coating to encapsulate the BZ media in. Disordered and ordered arrays of PE-coated BZ LMs were prepared to observe whether transfers of oxidation waves occurred between LMs in close proximity and to report propagation pathways within these different arrays.

PE-coated BZ LMs for the various arrays were prepared using \SI{50}{\micro\litre} and \SI{100}{\micro\litre} droplets of already oscillating BZ media. For disordered arrays of BZ LMs, a number of LMs of the same volume were rolled into a Petri dish. For the \SI{50}{\micro\litre} and \SI{100}{\micro\litre} BZ LM disordered arrays, the number of LMs used was 14 and 15 respectively. Unfortunately it proved difficult to completely fill the Petri dish, due to stability of the LMs.

\begin{figure}[!tbp]
\centering
    \includegraphics[width=0.3\textwidth]{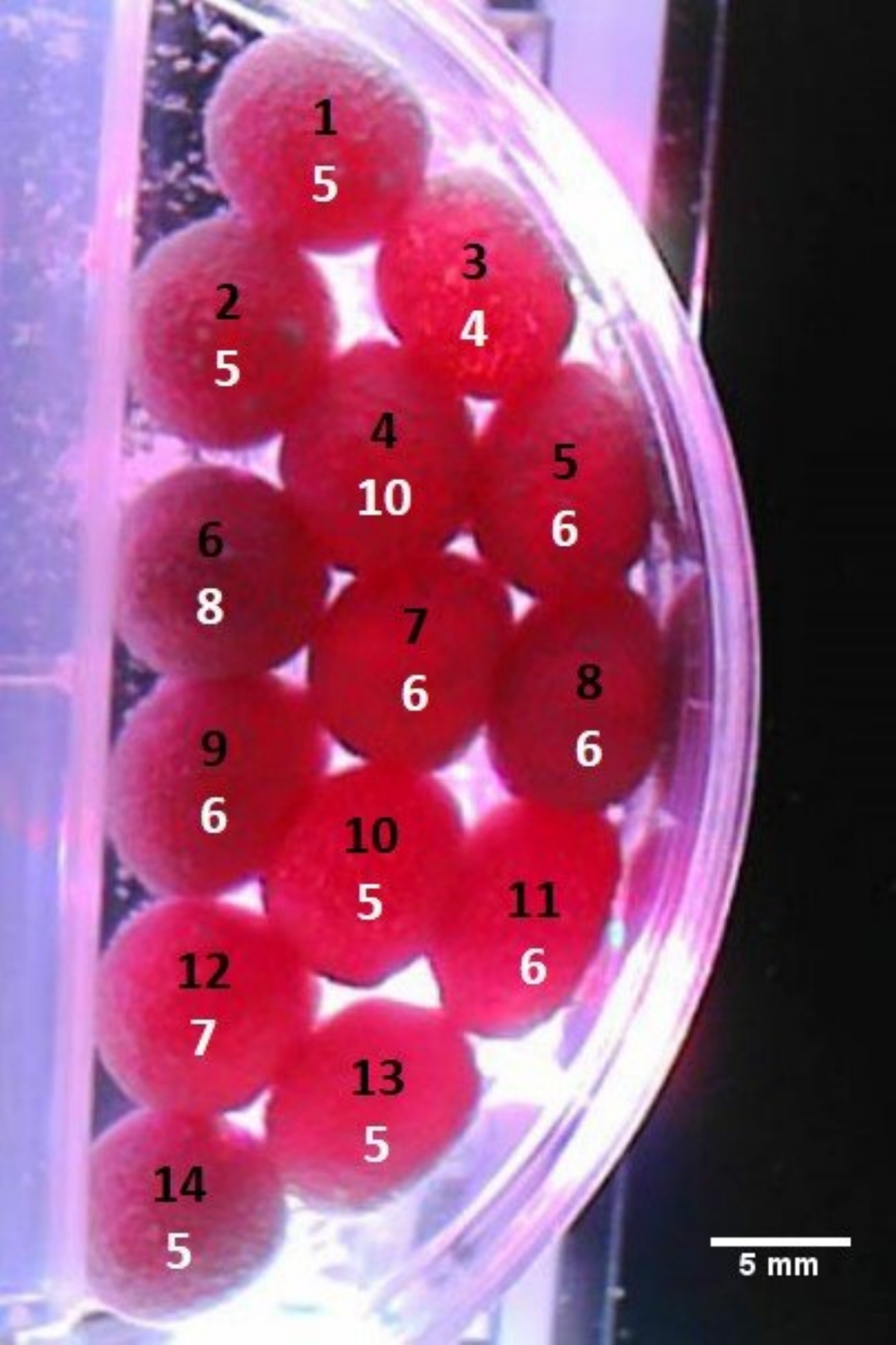} 
    \caption{\SI{50}{\micro\litre} PE-coated BZ LM disordered array -- top numbers (in black) refer to the numbers assigned to the LMs in the array, bottom numbers (in white) refer to the number of individual oscillations observed in each LM}
    \label{fig:disord50ul}
\end{figure}

\begin{figure}[!tbp]
\centering
    \includegraphics[width=0.4\textwidth]{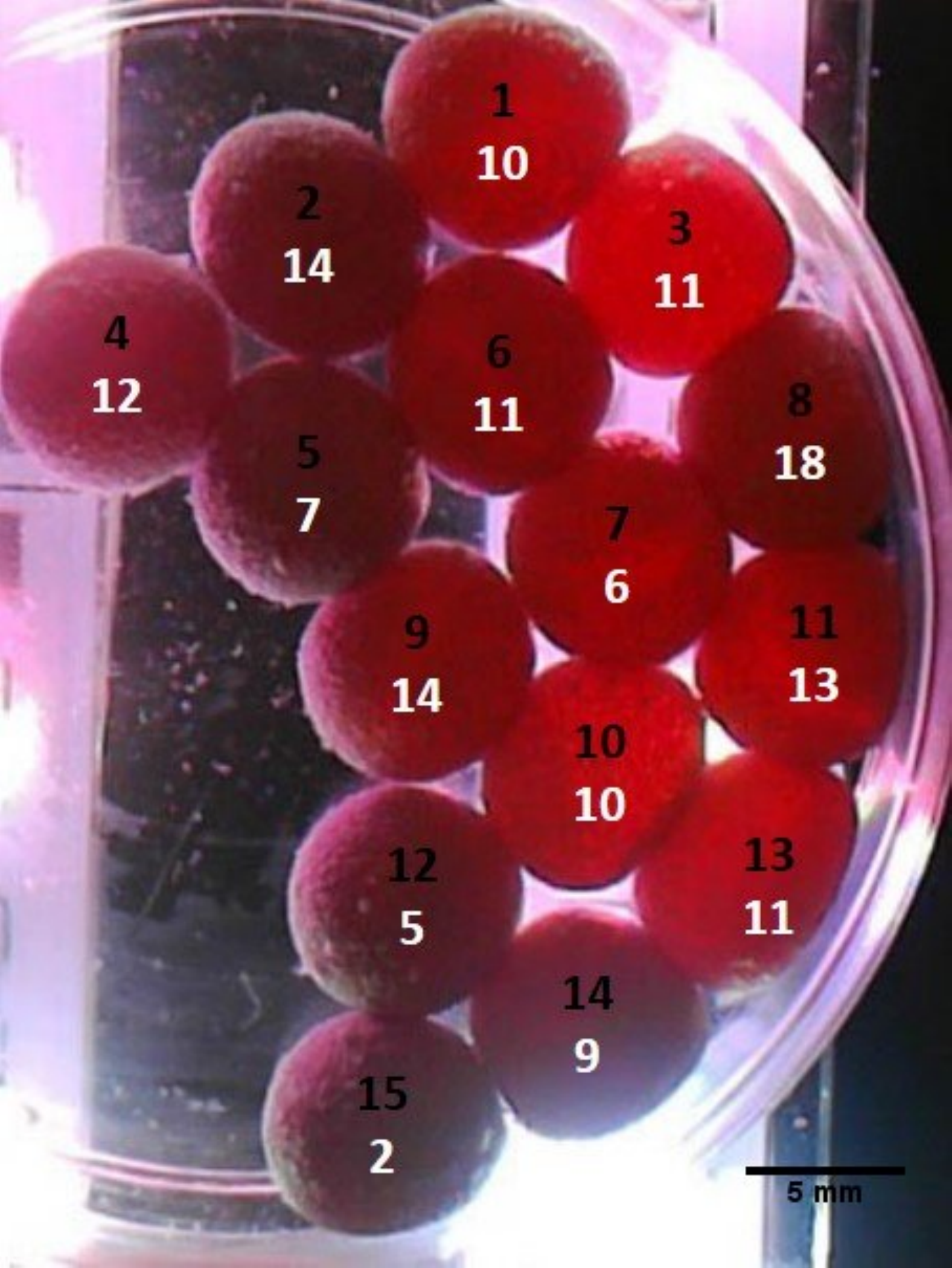}
    \caption{\SI{100}{\micro\litre} PE-coated BZ LM disordered array -- top numbers (in black) refer to the number assigned to the LMs in the array, bottom numbers (in white) refer to the number of individual oscillations observed in each LM}
    \label{fig:disord100ul}
\end{figure}

To discuss the transfer of waves and propagation pathways, BZ LMs in the disordered arrays were numbered, shown in Fig.~\ref{fig:disord50ul} and Fig.~\ref{fig:disord100ul} for \SI{50}{\micro\litre} and \SI{100}{\micro\litre} disordered arrays respectively. 84 individual oscillations were observed in the \SI{50}{\micro\litre} disordered array, with 14 of these waves resulting in transfer from one LM to another. Therefore, 17\% of oscillations resulted in transfers from one LM to another. All the LMs in the \SI{50}{\micro\litre} disordered array oscillated, LM4 oscillated the most with 10 visible wave-fronts, whilst LM3 oscillated the least with 4 visible wave-fronts. The number of individual oscillations each LM exhibited are reported in Fig.~\ref{fig:disord50ul}. Out of the 14 waves observed to transfer, 2 of these waves from 2 different LMs appeared to result in transfer to a single LM. The longest propagation pathway observed occurs between LM2 -- LM4 -- LM5 occurring from the 3rd and 4th wave transfers.

For the \SI{100}{\micro\litre} disordered array, 15 LMs were used, shown in Fig.~\ref{fig:disord100ul}. 153 oscillations were observed, 31 of which resulted in transfers from one marble to another. Therefore, 20\% of oscillations resulted in transfers from one marble to another, similar to the percentage observed for the \SI{50}{\micro\litre} disordered array. The number of individual oscillations each marble exhibited are reported in Fig.~\ref{fig:disord100ul}. The video for the BZ \SI{100}{\micro\litre} disordered array can be found in the ESI. The longest propagation pathway observed occurs between 3 LM marble in the \SI{100}{\micro\litre} disordered array, occurring from the 9th and 10th wave transfers.

\begin{figure*}[!tbp]
\centering
    \subfigure[820s]{\includegraphics[width=0.3\textwidth]{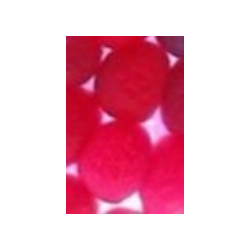}}
    \subfigure[825s]{\includegraphics[width=0.3\textwidth]{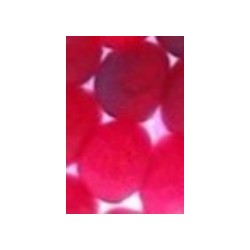}}
    \subfigure[830s]{\includegraphics[width=0.3\textwidth]{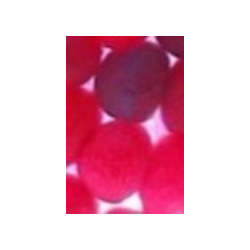}}
    \subfigure[835s]{\includegraphics[width=0.3\textwidth]{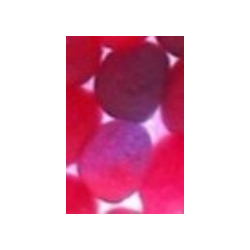}}
    \subfigure[840s]{\includegraphics[width=0.3\textwidth]{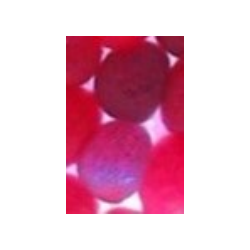}}
    \subfigure[845s]{\includegraphics[width=0.3\textwidth]{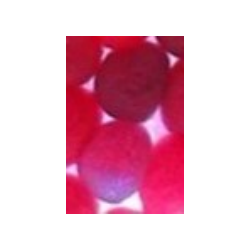}}
        \caption{1st transfer observed in \SI{50}{\micro\l} PE-coated BZ LM disordered array, transferring from LM7 to LM10}
    \label{fig:50ul_dis_t1}
\end{figure*}

As can be seen from \SI{50}{\micro\litre} disordered array transfer videos in the ESI and the 1st transfer from the \SI{50}{\micro\litre} disordered array shown as an example in Fig.~\ref{fig:50ul_dis_t1}, the direction of oxidation wave transfer between LMs demonstrates that excitation passes through the LM disordered array rather than being a result of spontaneous self-oscillation of the media inside a single LM. Similar wave transfers have been previously observed in BZ vesicles~\cite{DeLacyCostello2013}.

\begin{figure}[!tbp]
\centering
    \includegraphics[width=0.6\textwidth]{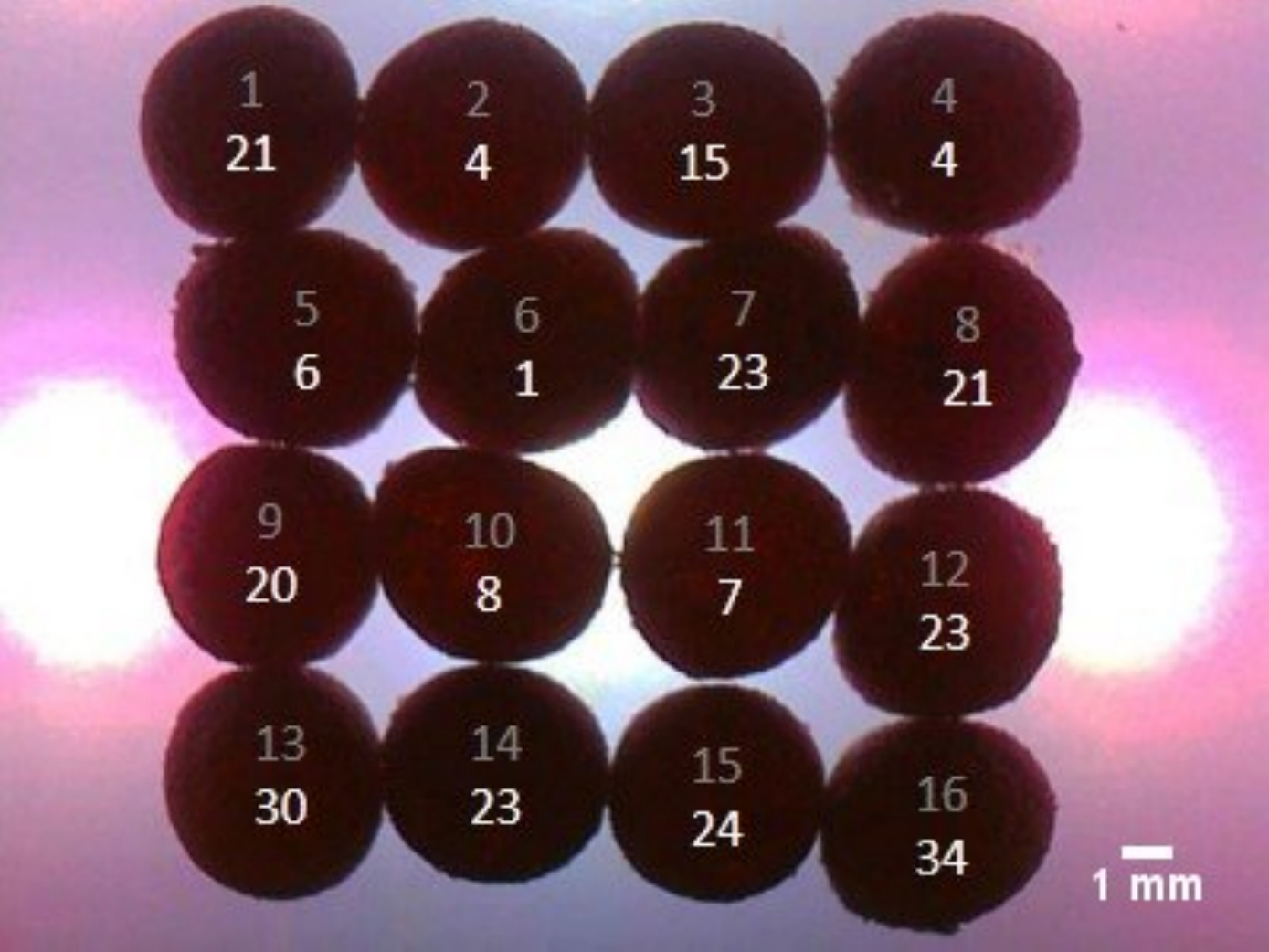}
    \caption{\SI{50}{\micro\litre} PE-coated BZ LM ordered array -- top numbers (in grey) refer to the numbers assigned to the LMs in the array, bottom numbers (in white) refer to the number of individual oscillations observed in each LM}
    \label{fig:16LM_50ul}
\end{figure}

Ordered arrays of \SI{50}{\micro\litre} BZ LMs were prepared by using a polypropylene template to position the marbles in a 4$\times$4 arrangement. This allowed more control over the number of contacts each LM had with adjacent LMs. Fig.~\ref{fig:16LM_50ul} shows a 16 LM ordered array recorded, in which 264 oscillations were observed. All LMs turned to the fully oxidised state after ca. 1 hour 20 mins. Fewer transfers occurred in the ordered array cf. the disordered array, with only 6 inter-marble transfers occurring, meaning only 2\% of oscillations transferred to an adjacent marble. The longest pathway observed involved 3 marbles (occurring during the 1st and 2nd transfers). In another 4$\times$4 ordered array, 182 oscillations were observed. All LMs turned to the fully oxidised state in this array after ca. 1 hour.

\section*{Discussion}

The BZ LM array experiments reported above show that it is feasible to observe oxidation waves through the coating of the LMs and observe transfers of these waves between adjacent LMs. It should be possible to control wave transfers by adjusting the chemistry of the encapsulated BZ system, for example using a different organic substrate or using a light sensitive catalyst, or by initiating and inhibiting the reaction to obtain controlled wave transfer. It is envisaged that through this controlled wave transfer and varying the arrangement of LMs in arrays, propagation pathways can be controlled and lengthened. This will allow the development of more complex computing devices using BZ LMs, in addition to observing the natural behaviour of the oscillating liquid media encapsulated in a powder coating, rather than in a liquid--liquid droplet system. The ability to compartmentalise the BZ media within a LM, which remains intact longer than the ferroin takes to oxidise, paves the way for future studies on BZ LM systems. Studying wave propagation in hexagonal arrays of BZ LMs, will provide a more natural arrangement for the LMs, in terms of spherical packaging and also advantageous in terms of increasing the number of neighbours for each LM within an array. This will aim to establish whether the number of interface contacts affects the oscillating nature of the BZ media inside LMs. Different techniques for controlled initiation of waves will be analysed, such as using a Ag wire, varying light intensity (if using the light sensitive catalyst) and / or altering the chemical environment of the LM e.g. adding methanol or formaldehyde to the BZ media inside the LM~\cite{vanRoekel2015} or tailoring the powder coating. In terms of tailoring the powder coating, it will be interesting to study the excitation dynamics in LMs made from BZ stock solution and impregnating the powder coating with BZ catalyst, as well as analysing the suitability of other powder coatings with catalysed BZ solution. It will also be interesting to explore methods for preparing the BZ reaction in-situ through marble merging~\cite{Draper2017, Liu2016a}, to compare with the results reported here for BZ LMs prepared by encapsulating the already oscillating BZ solution. Further evidence of liquid--liquid communication between LMs, as observed in the arrays reported above, will be sought through monitoring the transport of fluorescent particles through LM arrays. Diffusive coupling and excitation transfer has previously been observed in liquid based BZ systems, however in these cases the chemical communication is more established and intuitively expected cf. the novel LM system. Other methods of monitoring the BZ reaction within LMs will be assessed, such as tuning the setup of arrays and taking electrical measurements of the encapsulated BZ solution.

Numerical models of a light sensitive BZ medium encapsulated in geometric discs have demonstrated the feasibility of using this type of chemical system to implement polymorphic logic gates~\cite{Adamatzky2014}. The presence/ absence of oxidation wave-fronts at a given point were interpreted as {\sc True}/{\sc False} Boolean values with computations occurring via the interactions of these wave-fronts. By changing the illumination on the chemical media, it was possible to program the outcomes of computation between {\sc XNOR} and {\sc NOR} gates. Future studies will focus on implementing polymorphic logic gates with various arrangements of BZ LMs.

\section*{Conclusions}

Using LM preparation methods has proved to be a promising means of encapsulating the BZ media within small droplets that have solid--liquid interfaces. This work reports the first fabrication of BZ solution encapsulated in a powder coating, termed here as BZ LMs. The LM systems are novel in comparison to previous encapsulation methods e.g. BZ vesicles, as the inherent nature of LMs, enables studies into the behaviour of BZ solution in stirred and unstirred micro-reactor vessels, with the media confined within the LMs, and transitions between these two states. Oxidation waves were visible through the coatings of the LMs, so the reaction could be monitored. The versatility of LMs, in terms of being able to tailor the encapsulated liquid and powder coating provide an attractive system for encapsulating an oscillating media for the purposes of chemical information transmission and the development of new unconventional computing devices.

\section*{Conflict of interest}
There are no conflicts to declare.

\section*{Acknowledgements}

This research was supported by the EPSRC with grant EP/P016677/1 ``Computing with Liquid Marbles''.

\section*{Supporting Information}
Electronic supporting information is available. Videos listed below can be found on YouTube:

\begin{enumerate}
  \item PTFE-coated \SI{50}{\micro\litre} BZ LM -- \url{youtu.be/1U5e9ghmkqs}
  \item PTFE-coated \SI{100}{\micro\litre} BZ LM -- \url{youtu.be/bxfTlVjpBMU}
  \item PE-coated \SI{50}{\micro\litre} BZ LM -- \url{youtu.be/L4g8sk0v-t4}
  \item PE-coated \SI{100}{\micro\litre} BZ LM -- \url{youtu.be/dQtX-IQ_FXc}
  \item Disordered array of PE-coated \SI{50}{\micro\litre} BZ LMs -- \url{youtu.be/nO9CrYj1aPU}
  \item Disordered array of PE-coated \SI{100}{\micro\litre} BZ LMs -- \url{youtu.be/A30jNffrsQk}
  \item Ordered array of PE-coated \SI{50}{\micro\litre} BZ LMs -- \url{youtu.be/5zaroCTahjA}
\end{enumerate}

All videos were recorded at 1~fps, with playback at 10~fps.


\bibliographystyle{achemso3}
\bibliography{bibBZ}

\end{document}